\documentclass{aa}

\usepackage{graphicx}
\usepackage{lscape}
\newcommand\Msun{\rm\,M_\odot}

\usepackage{txfonts}
\usepackage{natbib}
\usepackage{makecell}
\usepackage[utf8]{inputenc}
\usepackage[T1]{fontenc}
\begin{document}

  \title{Massive star cluster formation and evolution in tidal dwarf galaxies}
\author{
 J\'er\'emy Fensch
         \inst{1,2}
          \and
          Pierre-Alain Duc \inst{3,2}
          \and
          M\'ed\'eric Boquien \inst{4}
          \and
          Debra M. Elmegreen \inst{5}
          \and
         Bruce G. Elmegreen \inst{6}
          \and
          Fr\'ed\'eric Bournaud \inst{2}
          \and
          Elias Brinks \inst{7}
          \and
         Richard de Grijs \inst{8,9,10}
          \and
          Federico Lelli \inst{1}
          \and
          Florent Renaud \inst{11}
         \and	
         Peter M. Weilbacher \inst{12} }

   \offprints{J. Fensch (jfensch@eso.org)}

   \institute{
   	 European Southern Observatory, Karl-Schwarzschild-Str. 2, D-85748 Garching, Germany
         \and
              AIM, CEA, CNRS, Universit\'e Paris-Saclay, Universit\'e Paris Diderot, Sorbonne Paris Cit\'e, F-91191 Gif-sur-Yvette, France
         \and 
              Universit\'e de Strasbourg, CNRS, Observatoire astronomique de Strasbourg, UMR 7550, F-67000 Strasbourg, France
           \and
              Centro de Astronom\'ia (CITEVA), Universidad de Antofagasta, Avenida Angamos 601, Antofagasta, Chile
           \and
              Vassar College, Dept. of Physics and Astronomy, Poughkeepsie, NY, USA
           \and 
             IBM Research Division, T.J. Watson Research Center, 1101 Kitchawan Road, Yorktown Heights, NY 10598 USA
           \and
           Centre for Astrophysics Research, School of Physics, Astronomy and Mathematics, University of Hertfordshire, Hatfield, Herts, AL10 9AB, UK
           \and
            Department of Physics and Astronomy, Macquarie University, Balaclava Road, Sydney, NSW 2109, Australia
            \and
             Research Centre for Astronomy, Astrophysics and Astrophotonics, Macquarie University, Balaclava Road, Sydney, NSW 2109, Australia
             \and
              International Space Science Institute--Beijing, 1 Nanertiao, Zhongguancun, Hai Dian District, Beijing 100190, China
             \and
             Lund Observatory, Department of Astronomy and Theoretical Physics, Box 43, SE-22100, Lund, Sweden
             \and
             Leibniz-Institut f\"ur Astrophysik (AIP), An der Sternwarte 16, 14482 Potsdam, Germany 
           }

   \date{Accepted June the 6th}

 
\abstract 
{The formation of globular clusters remains an open debate.  Dwarf starburst galaxies are efficient at forming young massive clusters with similar masses as globular clusters and may hold the key to understanding their formation.}
{We study star cluster formation in a tidal debris - including the vicinity of three tidal dwarf galaxies - in a massive gas dominated collisional ring around NGC~5291. These dwarfs have physical parameters which differ significantly from local starbursting dwarfs. They are gas-rich, highly turbulent, have a gas metallicity already enriched up to half-solar, and are expected to be free of dark matter. The aim is to study massive star cluster formation in this as yet unexplored type of environment.}
{We use imaging from the \emph{Hubble Space Telescope} using broadband filters covering the wavelength range from the near-ultraviolet to the near-infrared. We determine the masses and ages of the cluster candidates by using the spectral energy distribution-fitting code CIGALE, carefully considering age-extinction degeneracy effects on the estimation of the physical parameters. }
{ We find that the tidal dwarf galaxies in the ring of NGC 5291 are forming star clusters with an average efficiency of $\sim40\%$, comparable to blue compact dwarf galaxies. We also find massive star clusters for which the photometry suggests that they were formed at the very birth of the tidal dwarf galaxies and have survived for several hundred million years. Therefore our study shows that extended tidal dwarf galaxies and compact clusters may be formed simultaneously. In the specific case observed here, the young star clusters are not massive enough to survive for a Hubble time. However one may speculate that similar objects at higher redshift, with higher star formation rate, might form some of the long lived globular clusters.}
{}

\titlerunning{Star clusters in TDGs}


 \maketitle 
%

\section{Introduction}
\label{intro}

Globular clusters (GCs) are among the oldest stellar structures in the Universe. Their redshift of formation is estimated to be around $z = 2-6$ from their stellar population but their formation channels are still debated \citep[see reviews by][]{Forbes18, Renaud18}. In particular, the formation environment must be able to host very dense and massive gas clouds to allow the formation of these bound stellar clusters. Based on theoretical grounds, it has for instance been proposed that galaxy mergers at high redshift could be an important formation channel of current metal-rich GC populations \citep{Ashman92,Li14,Kim17}. Giant gas clumps in high-redshift, gas dominated galaxies could also host a favorable environment for GC formation \citep{Shapiro10, Kruijssen15}. The metal-poor part of the GC populations is proposed to be formed in the high-redshift highly turbulent gas-rich dwarf galaxies, such as the \emph{little blue dots}  seen at redshifts 0.5-4 in the Hubble Frontier Fields \citep{Elmegreen17}, which would be accreted, with their GC populations, onto more massive galaxies \citep{Cote98,Elmegreen12b,Tonini13,Renaud17}. Unfortunately, current instrumentation cannot probe the physical conditions of the GC birth environment at high redshifts, except for exceptional cases of strong gravitational lenses \citep{Vanzella17a, Vanzella17b,Bouwens17}. Observational studies of star cluster formation have thus mainly focused on favorable environments for massive star cluster formation in the Local Universe so far. 

Local dwarf galaxies are particularly interesting for the problem of GC formation. In starbursting dwarfs one may typically find young massive star clusters (YMCs) with masses above $10^{5}~\Msun$ and radii around 3~pc, that is, in the mass and size range of GCs \citep[see e.g.][and references therein]{deGrijs13,Hunter16}. Furthermore, local starbursting dwarf galaxies, such as the blue compact dwarf galaxies (BCDGs), can have up to 50\% of their current star formation rate (SFR) occurring in YMCs \citep[see e.g.][]{Adamo11} Finally, old evolved dwarf galaxies typically have a very large number of old GCs per unit luminosity, also called specific frequency\footnote{The specific frequency (S$_{\mathrm{N}}$) is the number of GCs per unit -15 absolute magnitude in the V-band ($M_V$): S$_{\mathrm{N}}$ = N$_{\mathrm{GC}} \times 10^{0.4(M_V + 15)}$.}: that is, much larger than for late-type galaxies and similar to massive early-type galaxies \citep[see e.g.][]{Lotz04, Peng08, Georgiev10}. Although the conditions of formation of present-day YMCs and old GCs are most certainly significantly different, these observations suggest that dwarf galaxies provide a favorable environment for both the formation and the survival of massive star clusters.

To extend the parameter space of dwarf galaxy environments, we present a study of massive star cluster formation and survival in dwarf galaxies which differ significantly from typical starbursting galaxies: tidal dwarf galaxies (TDGs). These galaxies are formed from gas and stars originating from the outskirts of a massive galaxy after a galaxy-galaxy interaction \citep[see review by][]{Duc99}. Because of this particular mode of formation, they are typically young, gas-dominated and are expected to be dark-matter free. Most importantly, their gas content is pre-enriched in metals and may already have a metallicity of one third to half solar. Thus, they deviate from the luminosity-metallicity diagram and have a significantly higher metallicity than starbursting dwarfs for a similar luminosity \citep[see e.g.][]{Weilbacher03}. Previous studies of the formation of tidal tails have shown that star clusters and TDGs may be able to form together in some cases \citep{Knierman03, Mullan11}. Possible examples of tidal dwarfs at redshifts of 0.5-1 are presented by \citet{Elmegreen07b}.  It should be noted that the merger frequency was much higher at high-redshift so that their contribution to the GC and TDG formation may have been more important. However, mergers at high-redshift seem to be much less efficient at triggering an enhancement of star formation \citep{Rodighiero11, Perret14, Lofthouse2017}, probably because of their high gas fraction \citep{Fensch17}. The formation of GCs and TDG in high-redshift galaxies still needs to be investigated.

The system studied in this paper is composed of young TDGs\footnote{These galaxies formed in collisional rather than in tidal debris. Even though they are thus not formally of tidal origin they are also found in the halo of a more massive galaxy and share the same physical properties as \emph{bona fide} TDGs. We will therefore also use the term TDGs for these galaxies.} located in a huge HI ring (M$_{\mathrm{HI}} > 10^{11} \Msun$, \citealp{Duc98}) expelled from the massive galaxy NGC~5291 (distance: 63.5~Mpc\footnote{The previous studies used this NED distance, assuming the following cosmological parameters: \emph{h} = 73, $\Omega_m$=   0.27, $\Omega_\Lambda$ =   0.73.}, distance modulus of 34.0~mag), most probably after an encounter with a bullet galaxy around 360~Myr ago \citep{Bournaud07}. This ring hosts four gravitationally bound objects with masses as high as $2 \times 10^{9}~\Msun$ \citep{Lelli15}, in the range of dwarf galaxies. These TDGs have a gas to stellar mass fraction of $\sim 50\%$ \citep{Bournaud07, Lelli15} and their spectral energy distribution (SED) is consistent with no stellar population older than 1~Gyr \citep{Boquien09}. Their material has been pre-enriched inside the host galaxy: they typically show half-solar metallicity  \citep{Duc98, Fensch16}.

This unique system has an extensive wavelength coverage: 21-cm HI line observations with the VLA \citep{Bournaud07}, molecular gas (\citealp{Braine01}, Lelli et al., in prep.), far-infrared with PACS and SPIRE on Herschel (Boquien et al., in prep.), mid-infrared with Spitzer \citep{Boquien07}, H${\alpha}$ with Fabry-Perot interferometry on the ESO 3.2m \citep{Bournaud04} and optical IFU with MUSE \citep{Fensch16} and far and near ultra-violet with GALEX \citep{Boquien07}. Radio and optical spectroscopy have shown the kinematical decoupling of the TDGs from the ring and their complex internal dynamics. MUSE has probed the variation of ionization processes throughout the most massive TDG of this system. However, none of the previously used instruments had the spatial resolution to investigate the TDG substructures and star cluster population. 

In this paper we present optical and near-IR imaging data from the \emph{Hubble Space Telescope} (\emph{HST}) which covers three of these TDGs. The pixel size in the optical is 0.04$\arcsec$, which corresponds to 12~pc at the distance of NGC~5291, and is small enough to allow us to distinguish the expected YMCs formed inside the dwarfs. We obtained broadband imaging covering a wavelength range from the near-UV to the near-IR. This allows us to derive the mass and age distributions of these TDGs' star cluster populations and study their formation and survival up to several hundred Myr in this particular environment. \\

We present the data acquisition and reduction in Section~\ref{Obs}. The star cluster selection and photometry measurements are presented in Section~\ref{Technics}. The derivation of their physical parameters (masses and ages) is described in Section~\ref{Derivation}. We discuss the cluster formation efficiency and cluster evolution in Section~\ref{Discussion} and conclude the paper in \ref{Conclusion}.


\section{Observation and Data Reduction}
\label{Obs}

\begin{figure*}
\centering{
\includegraphics[width=18.5cm]{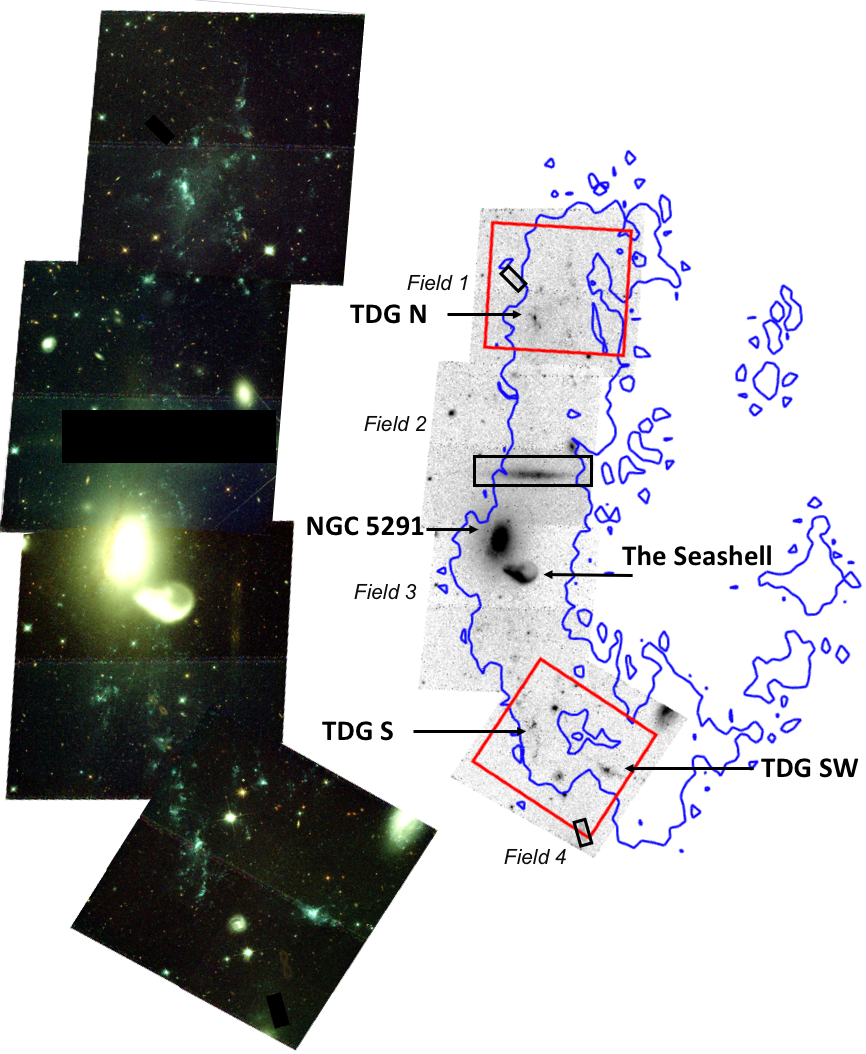}
\caption{Left: Composite color \emph{HST} image of the system using the F475W (blue), F606W(green) and F814W (red) filters.  North is up and East is to the left. Each field of view is 49.55~kpc $\times$ 53.34~kpc. Three regions contaminated by strong artefacts were masked. They are shown with black rectangles in the right image. Right: F475W image. The blue contour encircles regions where the HI column density is higher than $10^{20}$ N$_{\mathrm{HI}}$~cm$^{-2}$ \citep[VLA data, ][]{Bournaud07}. The two F160W-band fields of view are shown by the red rectangles. The central galaxy NGC~5291, the Seashell, and the three TDGs (N, S and SW) are indicated by black arrows, \label{map}}}
\end{figure*}

NGC~5291 collisional ring was observed with the WFC3 instrument on board the \emph{HST} (Project ID 14727, PI: Duc). The location of the field of view and the collisional ring are shown in Fig.~\ref{map}. We obtained photometry in the F336W, F475W, F606W, F814W and F160W bands. As will be discussed in Section~\ref{Derivation}, this set of filters was chosen for its ability to disentangle color effects from metallicity, age, and extinction in young star clusters \citep{Anders04}. The respective exposure times are given in Table~\ref{exptime}. We used the product of the regular MultiDrizzle reduction pipeline \citep{Koekemoer02}. The pixel size is 0.04$\arcsec$ for F336W, F475W, F606W, and F814W. The pixel size is 0.12$\arcsec$ for F160W. At the distance of NGC~5291 this corresponds to respectively 12~pc and 36~pc.\\

Only Field 1 and Field 4  were observed in the F336W and F160W bands. The field of view of the F160W data is slightly different, and is shown in the right part of Fig.~\ref{map}. The massive galaxy NGC~5291 and its companion (the Seashell Galaxy, \citealt{Duc99}) can be seen in the top part of Field 3, the TDG N in Field 1, and the the TDGs S and SW in Field 4. \\

The right-hand side image shows instrument artefacts: a bright saturation shape in Field 2 and the presence of 8-shaped reflection effects in Fields 1 and 4. Furthermore, one can see a small stripe of higher noise in the middle of each field of view: it is the location of the gap between the two UVIS CCDs of the camera, where we only have one exposure, and thus no cosmic ray removal. We allowed for any orientation to maximize the chance of observability. Unfortunately, the gap fell on both TDGs of Field 4. Only Fields 1 and 4 will be considered in the rest of the paper. Fields 2 and 3 will be the subject of a companion paper (Fensch et al., \emph{in prep.}).\\

We use the \emph{HST} image header keyword \textsc{PHOTFNU} to convert image units to Jansky. All magnitude values will be given according to the AB system in the following, unless specified otherwise.

\begin{table}[h!]
\centering
\caption{Exposure times for each field and filters in the following format: (number of exposures) x (time in second for a single exposure).\label{exptime}}

\begin{tabular}{l c c c c c}
 \\ \hline \hline 
                                                  & F336W                                              & F475W                                             & F606W   & F814W   & F160W           \\ \hline
Field 1  & 4 x 378.\ & 2 x 368.  & 2 x 368. & 2 x 368. & 4 x 903. \\
Field 2  & -                                                  & 2 x 368.                                           & 2 x 368. & 2 x 368. & -                                                  \\
Field 3  & -                                                  & 2 x 368.                                          & 2 x 368. & 2 x 368. & -                                                  \\
Field 4 & 4 x 378.  & 2 x 368.                                           & 2 x 368. & 2 x 368. & 4 x 903.\\ \hline 
\end{tabular}
\end{table}

\section{Cluster selection and photometry}
\label{Technics}

We extracted the star cluster candidates using SExtractor \citep{Bertin96} in the optical bands (F475W, F606W and F814W). The images were convolved through a \emph{mexican hat}-type filter\footnote{We used the filters provided in the SExtractor repository of astromatic.iap.fr. } with a width of two pixels to enhance the contrast with respect to the diffuse stellar light, and the detection threshold was set to 1.25~$\sigma$ for at least three adjacent pixels. 

As we only have two exposures for each of the three optical bands (F475W, F606W, F814W), the standard pipeline cannot remove cosmic rays which are coincident on the two exposures. We proceeded to apply a first cosmic ray subtraction by matching the location of the sources on these three filters. Only sources detected on at least both the F606W and either F475W and/or F814W images are considered for subsequent analysis. We also rejected 13 sources which are part of the GAIA DR2 \citep{GaiaDR2} catalog with a non-zero parallax and proper motion, which are likely foreground stars. After this step, we have 826 detections. This catalog of detections is then applied on the five bands to extract the photometry of the detected clusters.\\

The crowdedness of the sources in the TDGs prevented us from using a standard aperture photometry method. Instead, we used point spread function (PSF) fitting using GALFIT \citep{Peng02,Peng10b}. We first removed the background light using the sigma-clipping method implemented in SExtractor. In order to remove the diffuse stellar light in the TDGs we chose a tight mesh of 6x6 pixels, further smoothed with a 3x3 pixel kernel. The photometry was computed using PSF-fitting with GALFIT, using the PSF of the brightest unsaturated star available in the field. To avoid blending issues, we restricted the location of the peak of the PSF to vary by less than 0.08$\arcsec$, compared to the center of the detection in the F606W band. Some extracted sources appeared extended and were not well fitted in the F336W, F475W, F606W, and F814W bands. They were identified by a high pixel value dispersion in the residual image. The pixel size in these bands is 12~pc. In the early evolutionary stages of YMCs (1-10~Myr), the ionized gas surrounding the cluster may have a radius of around 20~pc, with a dependence on age \citep[see e.g.][]{Whitmore11}. One therefore expects to have barely resolved star clusters in these bands. For 40 sources out of the 826 detected sources, we performed S\'ersic photometry for proper subtraction.
To avoid unrealistic fits, we constrain the half-light radius to be smaller than 3~pixels and a S\'ersic index below 5. For consistency, we also fit the data without these constraints and the resulting values change only by less than half a standard deviation.
These sources were not resolved in the F160W image, which has a coarser resolution, hence we keep using PSF models for these sources. To ensure that the background subtraction method did not remove flux from either our point-like or extended sources, we verified the consistency within the error bars between the GALFIT method and an aperture photometry method for isolated sources. We used a 6~pixel radius and background estimation from the median background of 8 other same size apertures located around the source with an offset from the source of  -13, 0, +13 pixels for both the vertical and horizontal directions. Some sources were too faint for the GALFIT subtraction to converge. For those we used aperture photometry and considered this value an upper limit to the flux.

The error on the flux is obtained by combination of the Poissonian noise from the source and the removed background, the pixel-to-pixel root mean-square of the background, the pixel-to-pixel root mean square of the residuals, the systematic flux-dependent GALFIT flux uncertainty and the read noise from WFC3.\\

\begin{figure*}
\centering
\includegraphics[width=5.0cm]{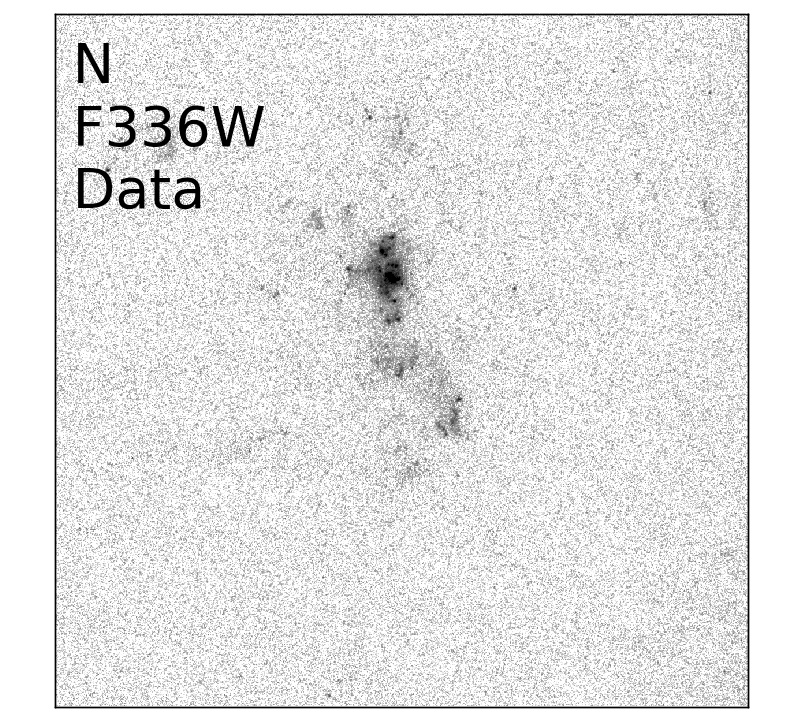}
\includegraphics[width=5.0cm]{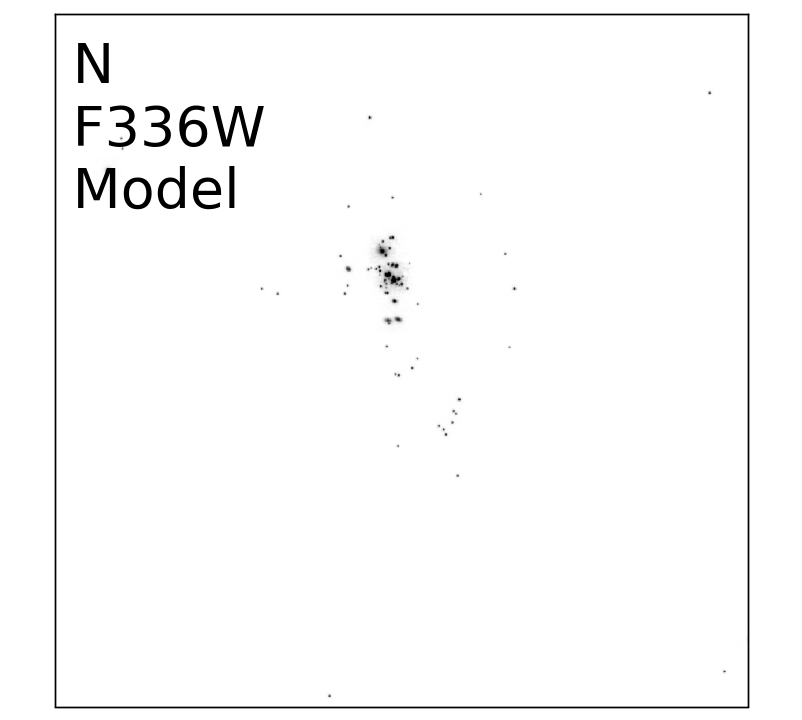}
\includegraphics[width=5.0cm]{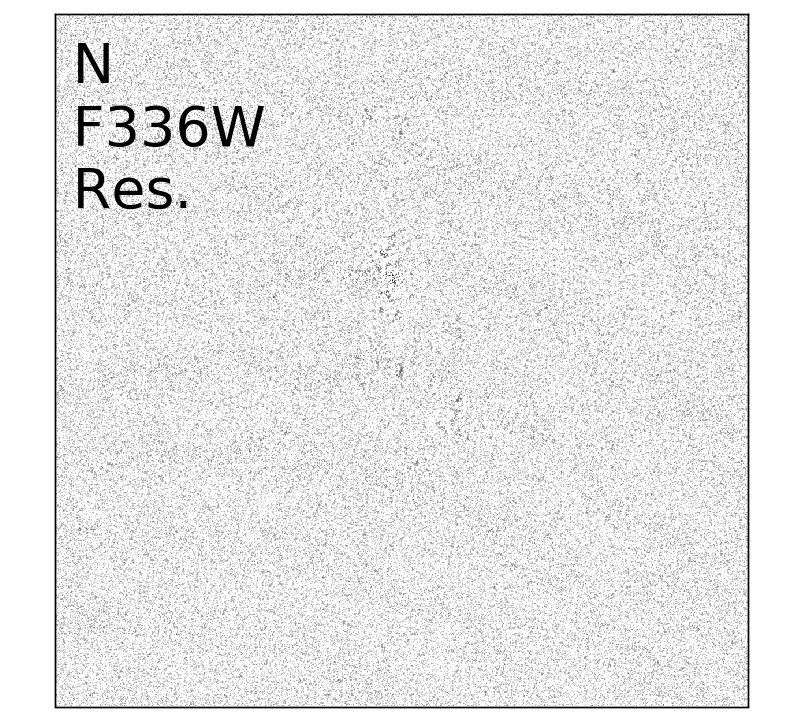}\\
\includegraphics[width=5.0cm]{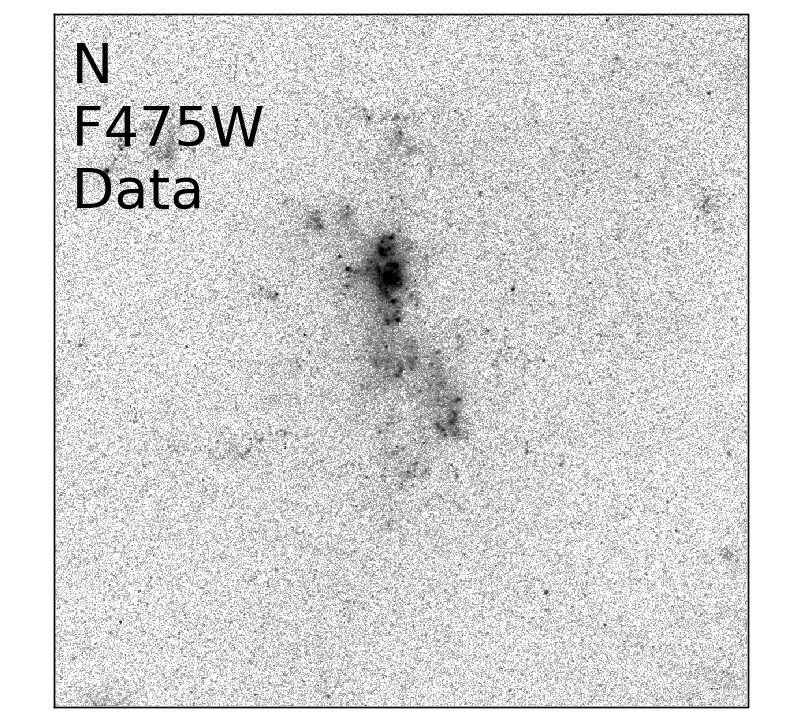}
\includegraphics[width=5.0cm]{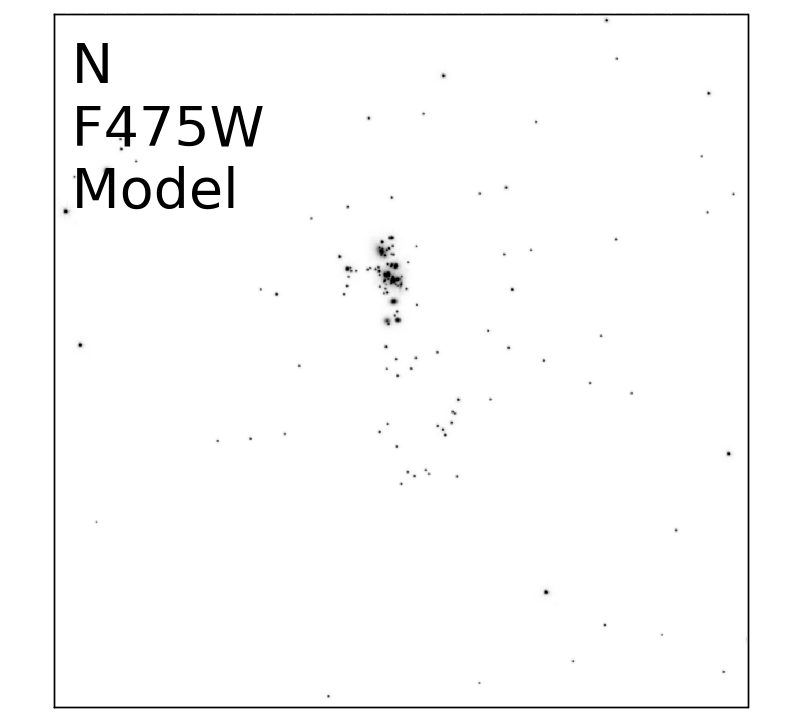}
\includegraphics[width=5.0cm]{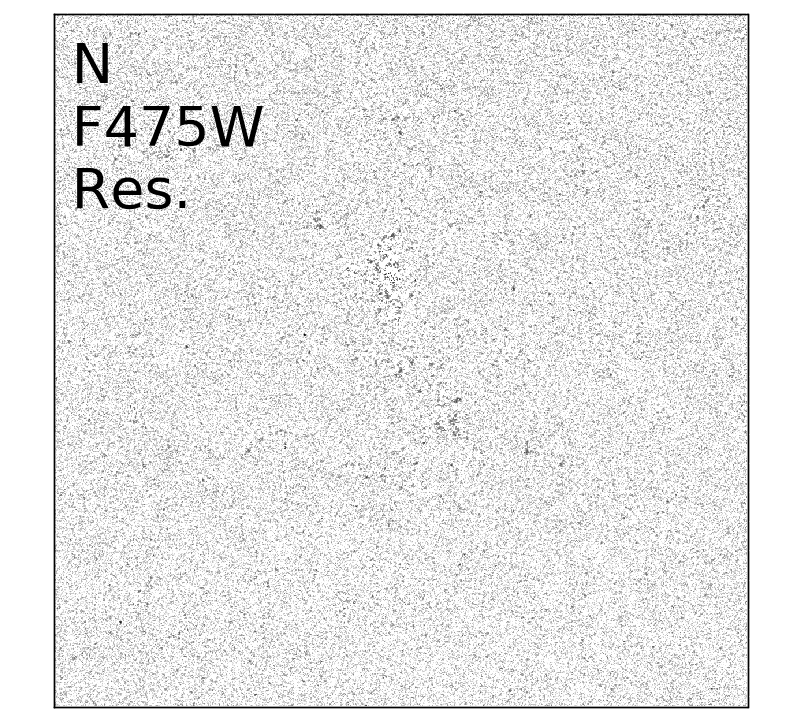}\\
\includegraphics[width=5.0cm]{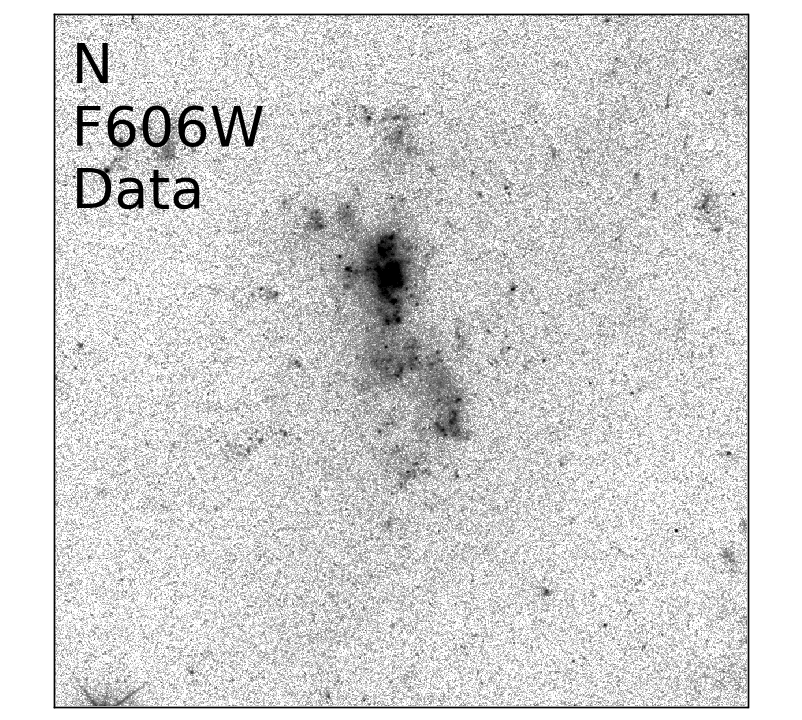}
\includegraphics[width=5.0cm]{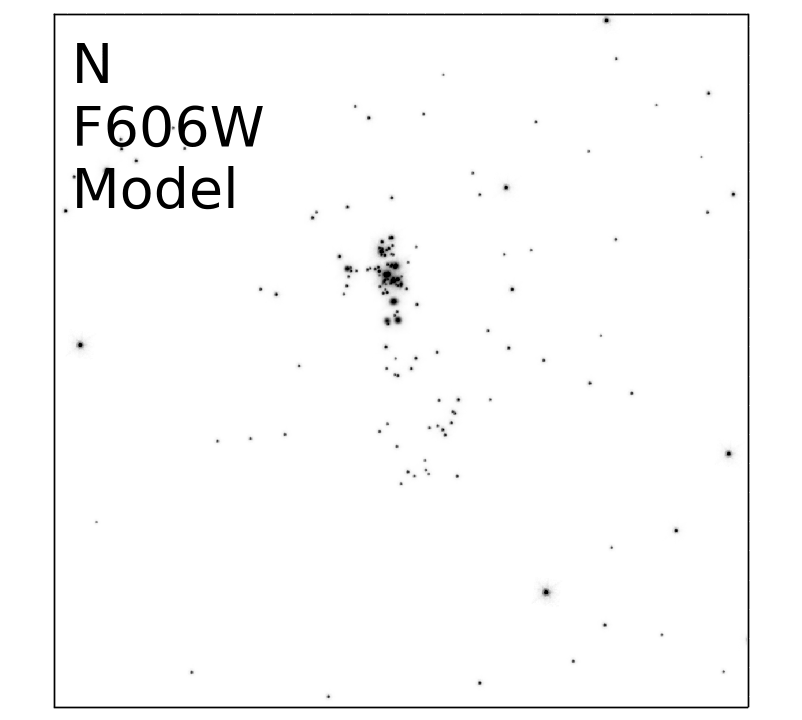}
\includegraphics[width=5.0cm]{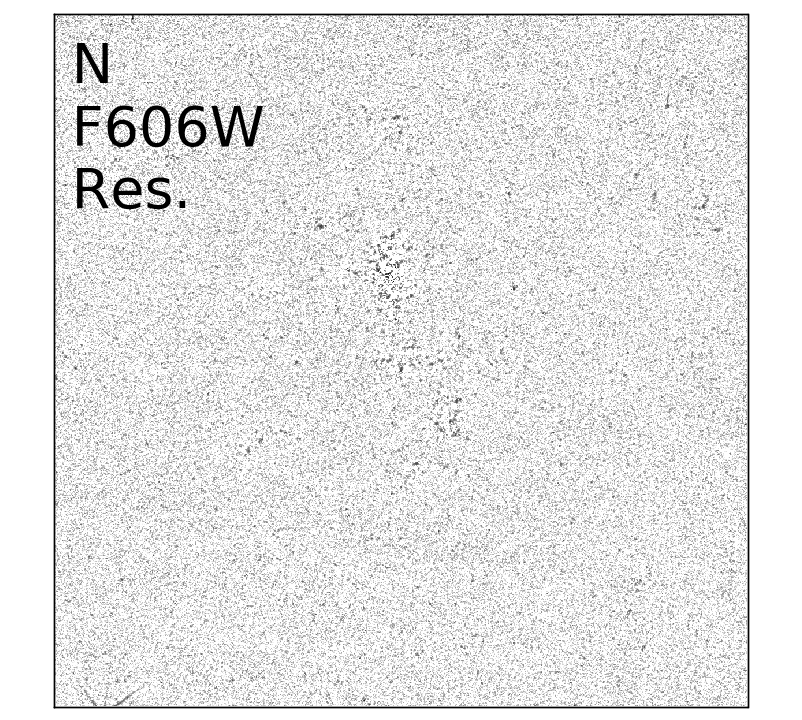}\\
\includegraphics[width=5.0cm]{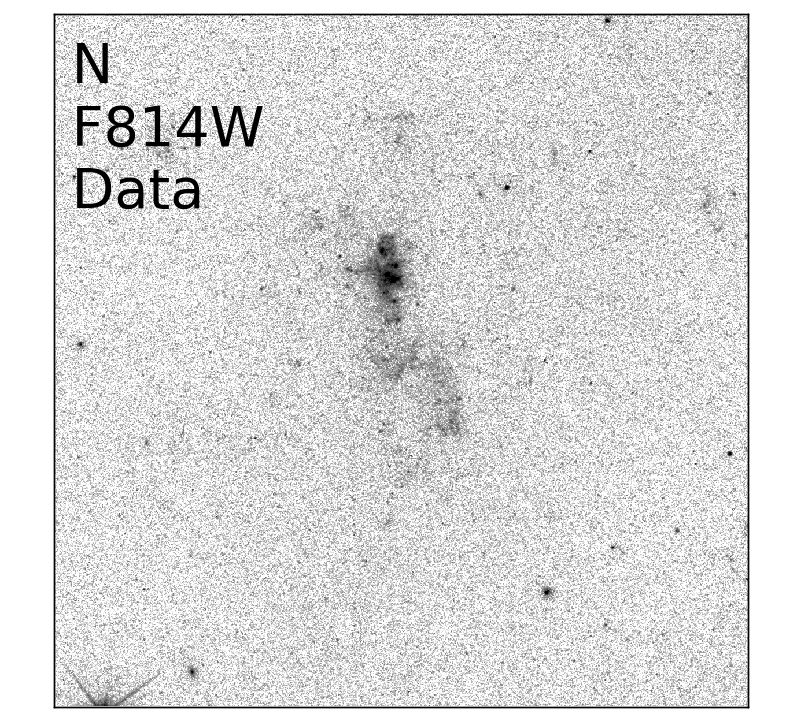}
\includegraphics[width=5.0cm]{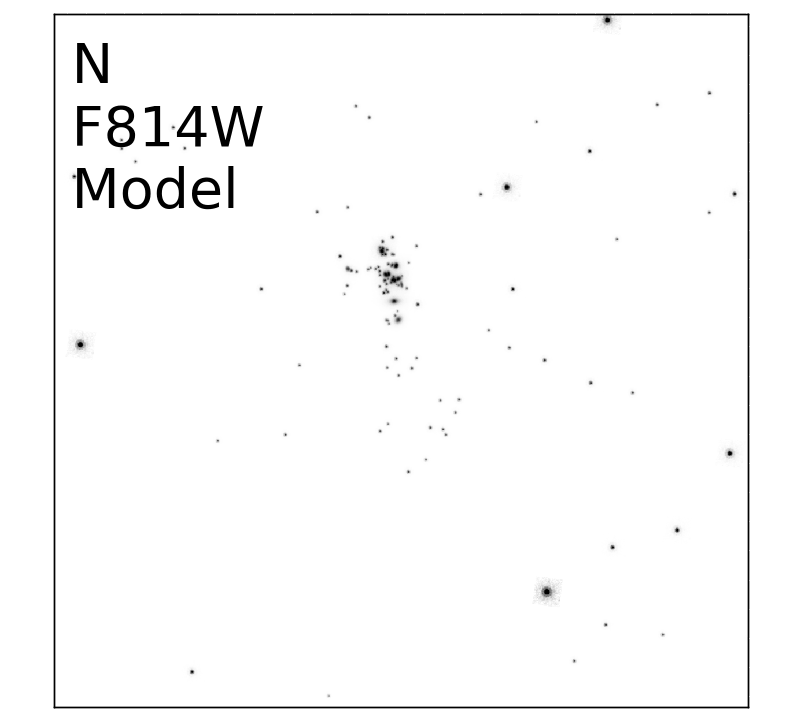}
\includegraphics[width=5.0cm]{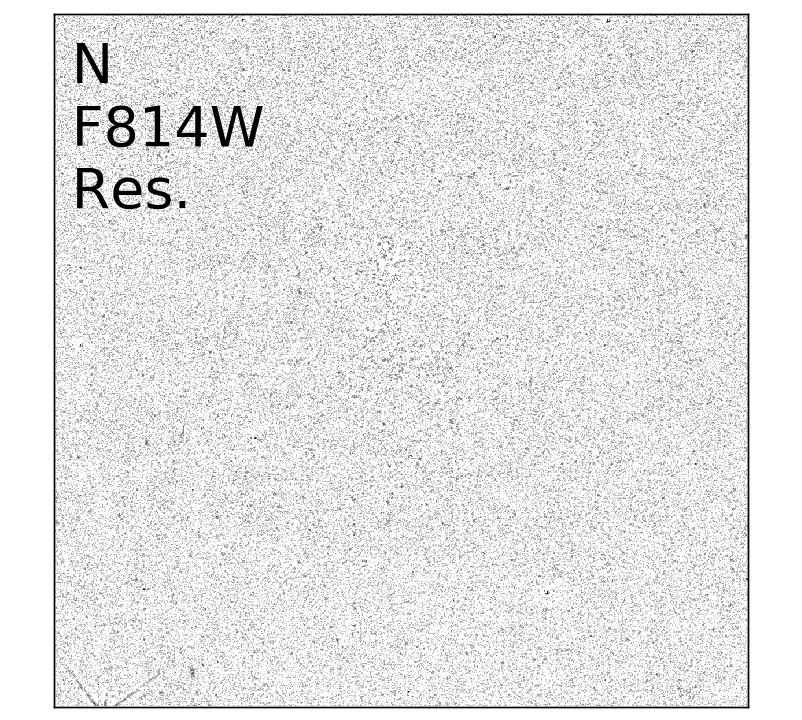}\\
\includegraphics[width=5.0cm]{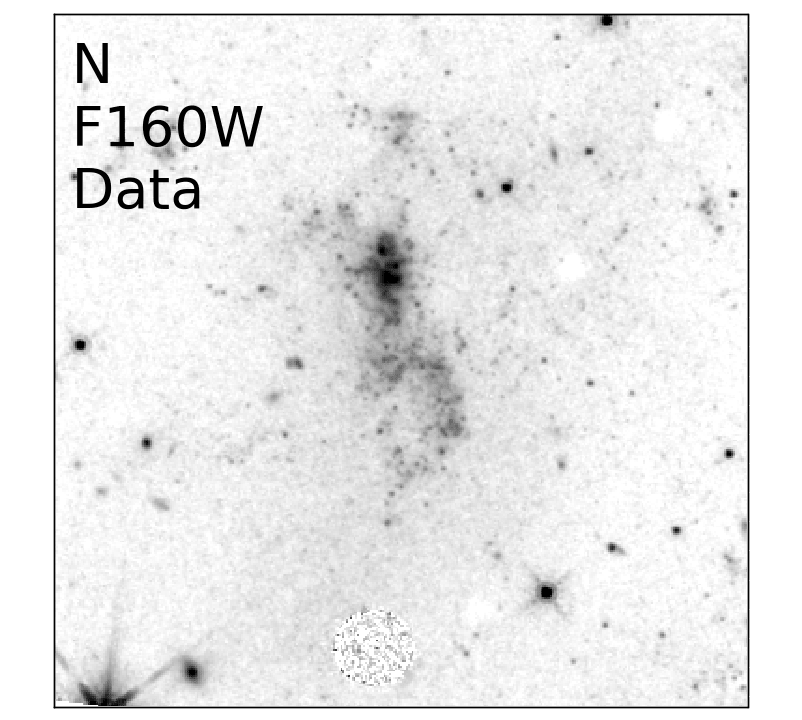}
\includegraphics[width=5.0cm]{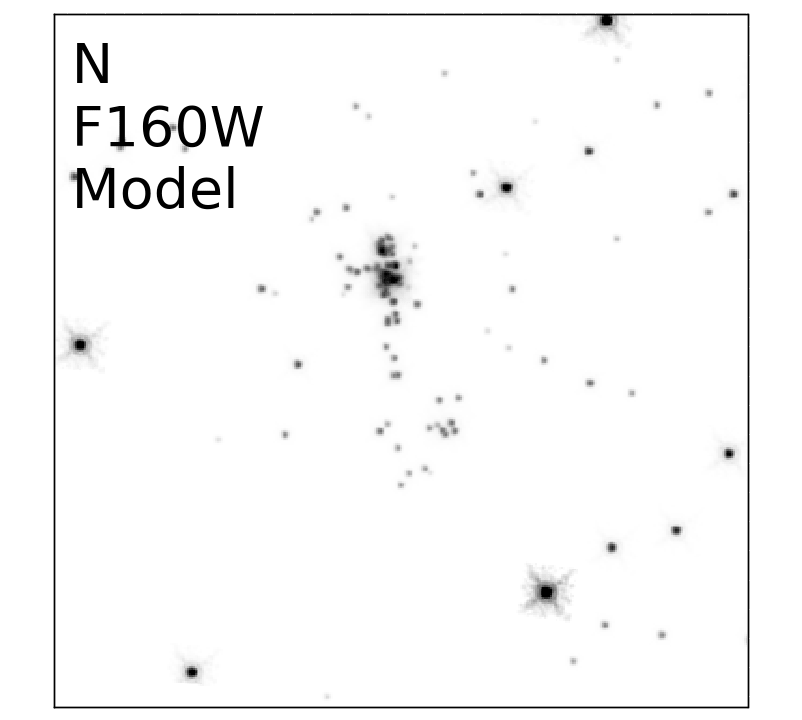}
\includegraphics[width=5.0cm]{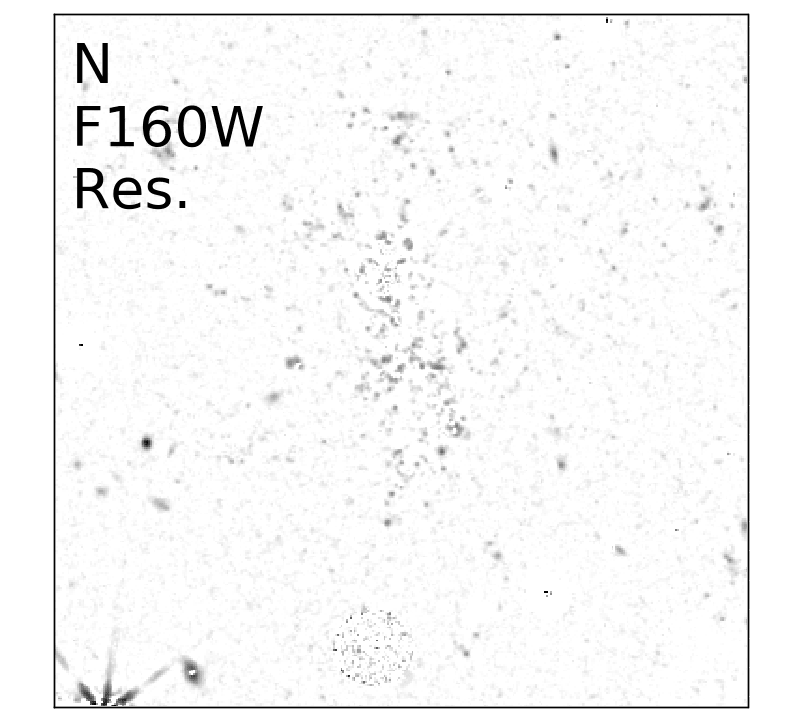}\\

\caption{Data, model and residual images for the TDG N. The two other TDGs are shown in Appendix~\ref{appendix::fits}. For each filter we show the data in the left column, the model of the clusters in the middle column and the background subtracted residuals in the right column. From top to bottom: F336W, F475W, F606W, F814W and F160W. We used the L.A. Cosmics algorithm \citep{vanDokkum01} to remove the cosmic rays. North is up and East is to the left.The field of view covers 14.4~kpc~x~14.4~kpc.\label{residual}}

\end{figure*}

A comparison of the original image and the residual after background and source subtraction is shown in Fig.~\ref{residual} for the three TDGs. In this figure some small extended stellar features remain on the location of star formation complexes. These features were not considered as detections by SExtractor, because of their elongated shape through the \emph{mexican hat} filter. They were also not accounted for by the background subtraction, being smaller than the background mesh. In our subsequent analysis we restrict ourselves to sources which have a signal-to-noise ratio higher than three in at least four bands, which leaves us with 439 cluster candidates. \\

\begin{figure}
\centering
\includegraphics[width=9cm]{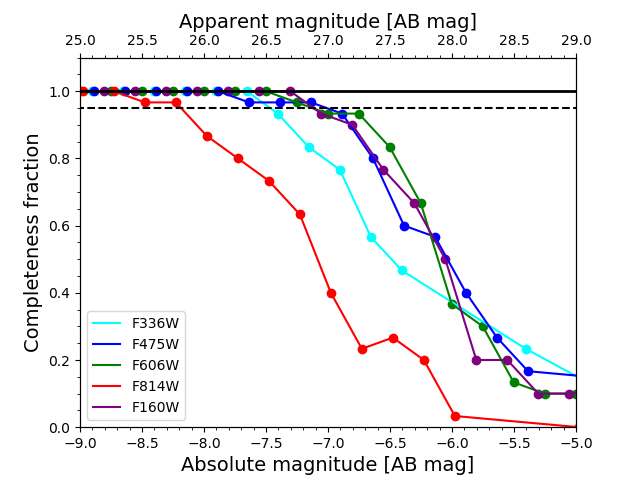}
\caption{Completeness curve of the star cluster detection algorithm for each filter. The horizontal dashed line at 0.95 shows the 95\% completeness limit. \label{complfig}}
\end{figure}

The completeness of our star cluster candidate extraction was computed by simulating point-like sources in the image using GALFIT and testing their detection and correct flux measurement using the same analysis as described above. A simulated source was considered recovered if detected by SExtractor and if its flux was recovered within 0.3~mag by GALFIT, as we accept a minimum signal-to-noise ratio down to 3 in a given filter. The completeness curves are shown in Fig.~\ref{complfig}. The 95\% completeness limit will be considered in the following.\\

\section{Deriving cluster physical properties}
\label{Derivation}

\subsection{SED fitting procedure}
\label{CIGALE}

The set of filters we used was chosen for its ability to recover ages and extinction for young star clusters using SED fitting procedures \citep{Anders04}. In this work, we used the SED fitting code CIGALE\footnote{Code available at https://cigale.lam.fr} \citep{Burgarella05, Noll09, Giovannoli11, Boquien19}, This code first compu.es a grid of flux models for a given input of discrete parameters from the stellar models of \citet{Bruzual03}, normalized to a fixed mass. In a second step, the code performs a $\chi^{2}$ analysis between the source and the flux grid, including a normalization to obtain the mass corresponding to the fit. \\

We have chosen the following range of parameters:
\begin{itemize}
\item{{\bf Star formation history:}  We use the \citet{Chabrier03} initial stellar mass function with lower and upper mass limits of respectively 0.1 and 100~$\Msun$. We model our clusters as a single quasi-instantaneous burst of star formation with an exponential decay with a 0.1~Myr timescale. To quantify the senstivity of the results on this parameter, we also modelled the star formation burst with a 1~Myr timescale. The small variations in the resulting values are quantified in the following and do not affect our conclusions.} 
\item{{\bf Age:} to account for both very young star clusters as well as GCs, we used models from 1~Myr to 12~Gyr. We use an adaptive spacing to account for the rapid change of the spectra at young ages. In particular, we have one model per Myr from 1 to 20~Myr and one per 5~Myr from 20 to 50~Myr. The weights of the fits depend on the age grid spacing, in order to have a flat age prior.}
\item{{\bf Metallicity:} the metallicity of the ring is approximately constant at around half solar metallicity \citep{Duc98, Fensch16}. We therefore fix the metallicity to Z = 0.008 to avoid degeneracies with age and extinction. The impact of changing the metallicity prior will be discussed in the following sections.}
\item{{\bf Extinction:} we use the LMC extinction curve from \citet{Gordon03} as it is the most suitable to our half-solar metallicity. The extinction obtained from MUSE from the Balmer decrement and the LMC extinction curve gave extinction values of the order of A$_{V} = 0.6 \pm 0.2$ mag throughout the northern TDG, on a spatial scale of 180~pc~x~180~pc \citep{Fensch16}. NASA infrared science archive service\footnote{https://irsa.ipac.caltech.edu/applications/DUST/} indicates a Milky Way extinction value in the line-of-sight of NGC~5291 of around 0.15 mag. In order to stay conservative, we allow for extinction ranging from A$_{V}$ = 0 to 2~mag.}
\end{itemize}

Furthermore, we allow the ionisation parameter log U to vary between -4 to -2 with 0.5 dex steps, according to the range determined from emission line ratios with MUSE \citep[see Fig.10 in ][]{Fensch16}. Finally we allow for a fraction of escaping Lyman continuum photons between 0 and 20\% \citep[dwarfs with strong outflows have a fraction of escaping Lyman continuum photons around 15\% see e.g. ][]{Bik15}. We assume a gas density of 100~cm$^{-3}$ \citep{Fensch16}.

\subsection{Physical parameters and degeneracies}
\label{sub_deg}

\begin{figure*}[h!]
\centering 
\includegraphics[width=6.5cm]{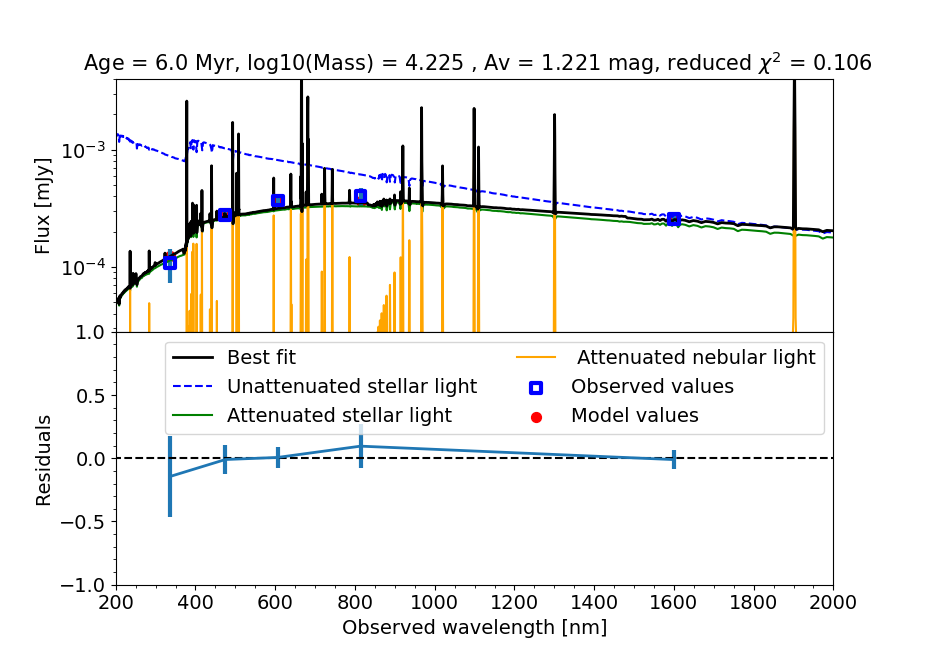}
\includegraphics[width=6.5cm]{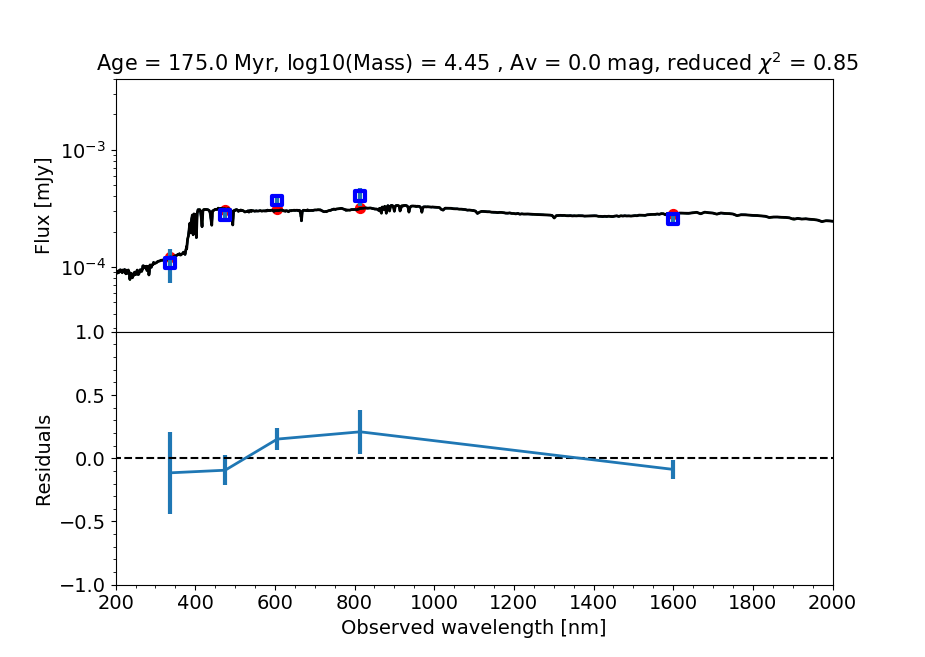}\\
\includegraphics[width=6.5cm]{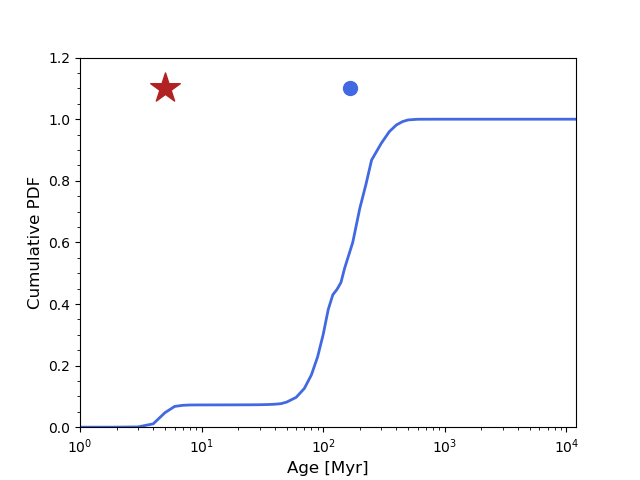}
\includegraphics[width=6.5cm]{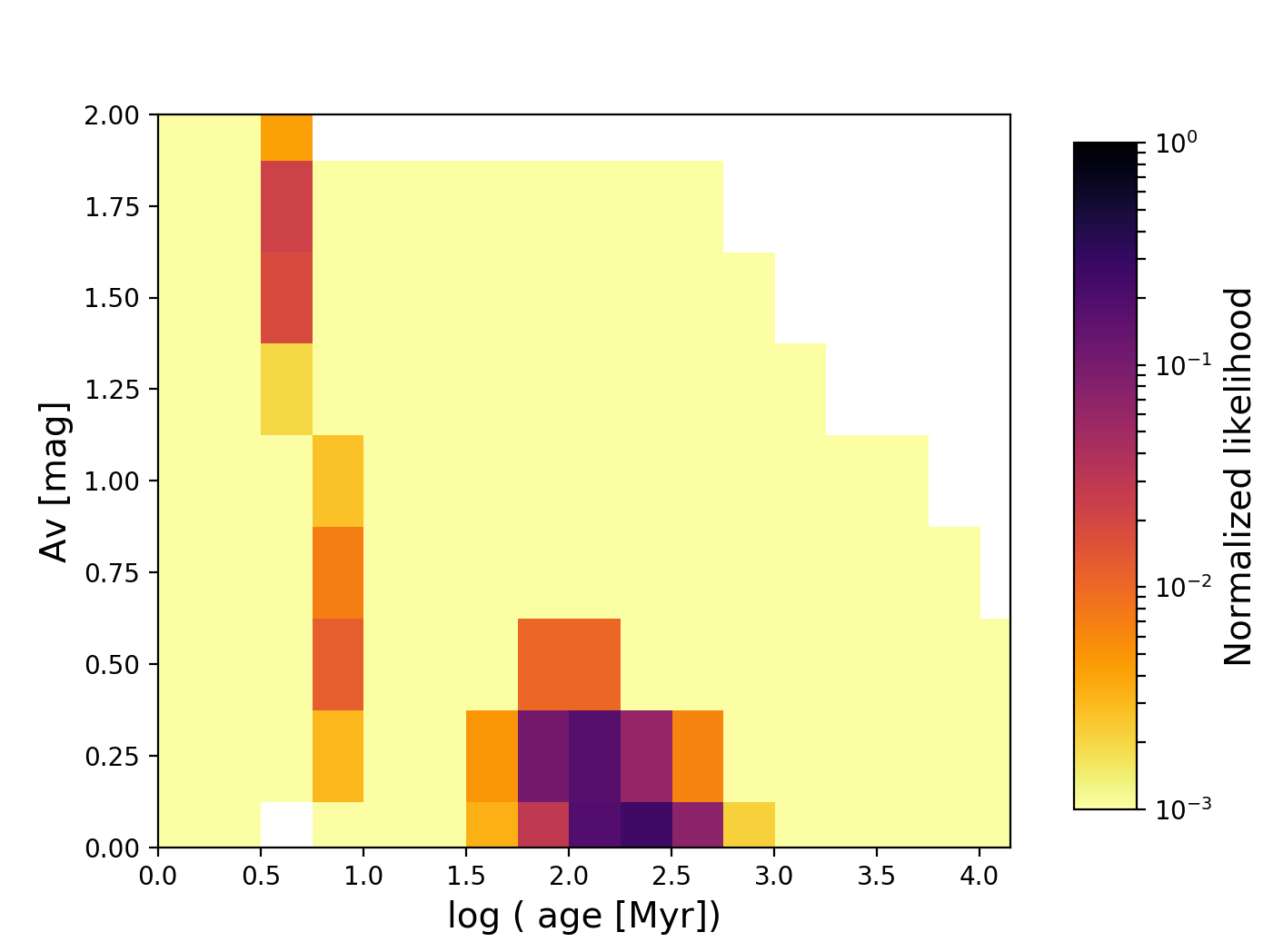}
\caption{Top left: example of the best fit for one cluster candidate. The retrieved physical parameters and the reduced $\chi^{2}$ are given in the title of the plot. Top right: Best fit for the same cluster candidate if one imposes A$_V = 0$~mag. Bottom left: cumulative age PDF for the cluster. The star shows the best fitting age. The blue dot shows the output value of CIGALE. Bottom right: Normalized likelihood distribution for age and extinction for the given cluster candidate.\label{deg}}
\end{figure*}

We are interested in recovering good estimates of the ages and masses of the clusters. However, for our set of filters, there is a degeneracy between extinction and age, which is illustrated in  Fig.~\ref{deg}. In the top panels one can see two models which fit the data well, with very different ages and extinction values. The cumulative probability density function shown in the bottom-left panel shows two characteristic values for the age and its rise is quite extended. The origin of this wide distribution is an age-extinction degeneracy: in the bottom right panel we see that both young and attenuated models, and old and unattenuated models can reproduce our photometry for this particular cluster. Even though their metallicity is known, this degeneracy prevents us from deriving precise ages for all clusters.

To quantify this effect, we use a proxy for the width of the age PDF: $r_{\mathrm{age}} = \mathrm{F}_{0.95} / \mathrm{F}_{0.05}$, where $F_ {x}$ is the age for which the cumulative age probability distribution function reaches $x$. The extent between $\mathrm{F}_{0.95}$ and $\mathrm{F}_{0.05}$ is shown in the bottom left panel of Fig.~\ref{deg}. The particular cluster candidate shown in Fig.~\ref{deg} has  $r_{\mathrm{age}} = 58$. We note that emission-line information, such as H$\alpha$ emission mapping at the scale of the size of a star cluster (10-20~pc) would help break this degeneracy: the presence of ionized gas would classify a given cluster as unambigously young \citep[see e.g.][]{deGrijs13}. However, we cannot distinguish the star clusters on the emission-line maps we have obtained with MUSE and Fabry-Perot interferometry (see Section~\ref{intro}).\\

\begin{figure}
\centering
\includegraphics[width=10cm]{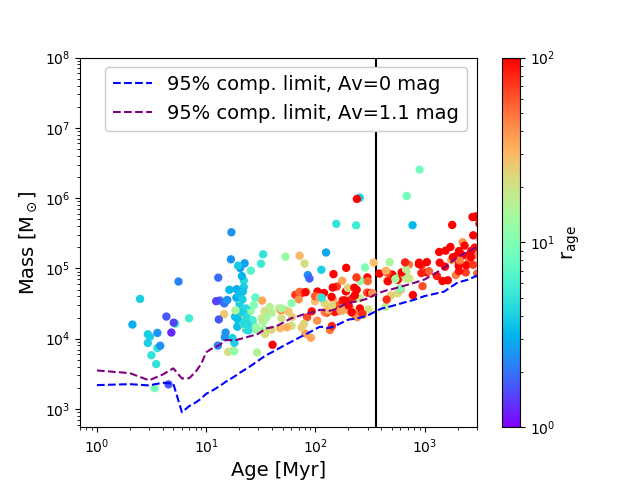}
\caption{Estimated mass-age distribution for the cluster candidates. The color indicates the width of the age PDF, as defined in the text. The two dashed lines indicates the 95\% completeness limit in the age-mass plane, assuming a given extinction. The black line shows the estimated time of the interaction which created this system (see text). \label{mass_age}}
\end{figure}

Fig.~\ref{mass_age} shows the retrieved masses and ages for the cluster candidates, where the width of their age PDF is color-coded. We note an increase of $r_{\mathrm{age}}$ with age, which is due to the slower spectral evolution with age, and also a significant number sources for which $r_{\mathrm{age}} > 50$ which might be subject to degeneracies. In this study we are not interested in old GCs. This is the reason why we only consider clusters with retrieved age below 3~Gyr in Fig.~\ref{mass_age}. Older clusters, for which the fixed half-solar metallicity prior is not adapted for mass and age determination, will be presented in a companion paper, along with the study of Fields 2 and 3 (Fensch et al., \emph{in prep.}).\\

Fig.~\ref{mass_age} also shows the completeness limit as obtained in Section~\ref{Technics}. These curves were obtained from the flux models computed by CIGALE. For each age, the curves show the minimum mass for which a cluster would have a 95\% probability to be detected with a S/N ratio above 3 in at least 4 bands. We show the completeness curves for two assumed extinctions, A$_{\mathrm{V}} $ = 0 and 1.1~mag. This latter value was the maximum extinction obtained using the Balmer decrement from the field of the Northern TDG in \citet{Fensch16}. Based on this figure, we assume that our sample is complete for clusters younger than 30~Myr above a mass of $1.5\times10^4~\Msun$.

\subsection{Young cluster mass function}
\label{young}

In the following we explore the properties of the young clusters. In order to have a sufficient number of detections, we chose to consider only clusters with an age below 30~Myr. 

Given the degeneracy effect, one cannot compile a complete sample. Indeed, young and strongly attenuated clusters could in principle masquerade as old and unattenuated clusters. To define our sample, we will use the shape of the age PDF.  In particular, we write 

\begin{equation}
\mathrm{P}[\mathrm{age} < X] = \int_{0}^{X} \mathrm{PDF}(t) ~ \mathrm{d}t 
\end{equation}

with X in Myr, the probability that the cluster candidate has an age younger than X~Myr.\\

In the following, we define our young cluster sample adopting  $\mathrm{P}[\mathrm{age} < 40]  >  0.5$ and the modal value of the age PDF being enclosed in [1~Myr, 30~Myr].\\

We used 40~Myr as the upper bound for the PDF integral because using 30~Myr led to rejecting clusters with ages between 20 and 30~Myr, which have a large fraction of their PDFs extending beyond 30~Myr. We then chose a higher upper bound for the integral calculation, and use the condition on the mode of the PDF to ensure that the highest likelihood is still reached within the [0,30] Myr interval. To quantify this effect, we use this sample selection in association with the spectral models created by CIGALE with input ages between 20 and 40~Myr. Only 28\% of the clusters with input age in the range [20,30]~Myr are included in our sample if we use 30~Myr as upper bound, while this fraction rises to 71\% if one uses 40~Myr as upper bound. The contamination fraction, that is the fraction of models assigned to the sample which have ages in (30,40]~Myr, is 0\% in the first case and 6\% in the second case. Using 40~Myr instead of 30~Myr in the definition of our sample therefore gives a better representation of the clusters with ages genuinely younger than 30~Myr. \\

We discuss in Appendix~\ref{Annex_deg} two other sample selections: a \emph{Secure} sample, defined by P[age < 40] > 0.9, and an \emph{Inclusive} sample defined by P[age < 40] > 0.1. The exact same analysis is performed on these two samples for comparison purposes. Since the former is very restrictive and the latter will include clusters that are too old, this additional analysis gives an idea of the strict boundaries within which our result may vary.

\begin{figure}
\centering
\includegraphics[width=10cm]{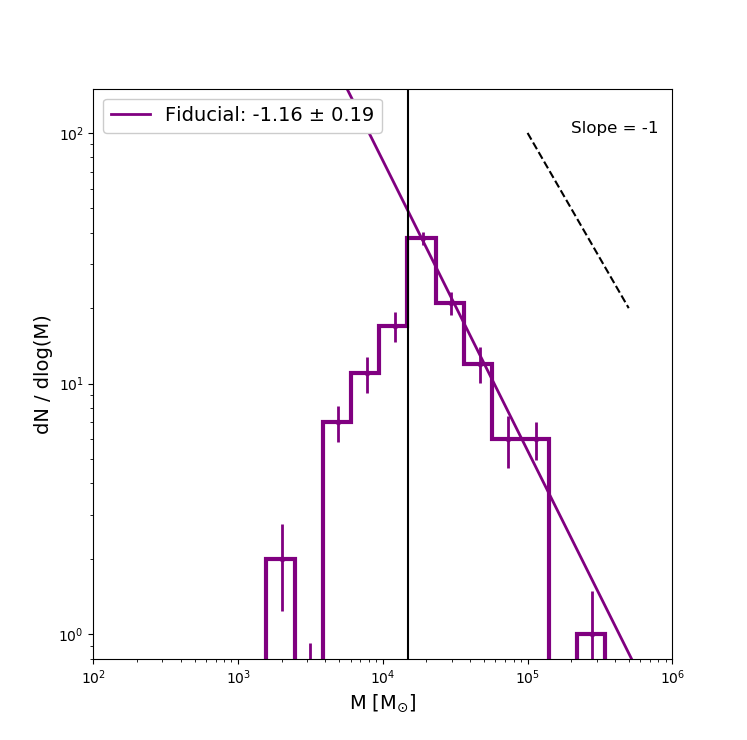}
\caption{CMFs for the young cluster sample described in the text. The power-law fit is determined for bins with masses higher than $1.5\times10^4~\Msun$, shown by the black vertical line. The legend shows the slope and uncertainty of the corresponding fit. \label{CMF}}
\end{figure}

Fig.~\ref{CMF} shows the cluster mass function (CMF) for our young cluster sample.  A power-law fit to the diagram, for bins more massive than $1.5\times10^4~\Msun$, gives a slope of $-1.16\pm0.19$ for the evolution of  $\frac{dN}{dlogM}$ with $M$. This gives $\frac{dN}{dM} = \frac{dN}{M dlogM} \propto M^{\alpha}$, with $\alpha = -2.16 \pm 0.19$. The values obtained for the lower metallicity prior (Z=0.004) and for the 1~Myr timescale are consistent within the 1-sigma uncertainty. The obtained mass distribution is consistent with a mass distribution decreasing with a power-slope of $\alpha \sim -2$, as in many other studies of young star cluster formation have shown \citep[see e.g.][]{Portegies10}. This suggests that the formation of star clusters in the gas ring and TDGs occurs in a similar fashion to that of the other studied environments. This can be interpreted as a legacy of the hierarchical collapse of gas clouds \citep[see e.g.][]{Elmegreen97}. 

\subsection{Star cluster formation efficiency in the TDGs}
\label{res_gamma}

\begin{table}[]
\centering{
\caption{SFR, Area and CFE for the presented TDGs. The last two columns show the expectation of the CFE from the \citet{Kruijssen12} and \citet{Johnson16} models for the measured SFR surface density, named respectively K12 and J16.} 
\label{cfe_fid}
\begin{tabular}{l c c c c c }
\hline \hline
                          & SFR                 &  Area          & CFE   &  K12 & J16  \\
Galaxy                            &   [$\Msun$/yr]                 &   [kpc$^{2}$]            &   [$\%$]   &   [$\%$]   &   [$\%$]        \\ \hline 
TDG N                     & $ 0.19 \pm 0.06                                  $ &$ 12.76     $        &       47$^{+21}_{-21}$    & $14^{+2}_{-2} $ &  $22^{+10}_{-9} $          \\ 
TDG SW                   & $ 0.14 \pm 0.05                           $      & $ 17.44           $ &      33$^{+17}_{-16}$     & $10^{+2}_{-2} $ &    $15^{+8}_{-7} $            \\ 
TDG S  &                       $ 0.12 \pm 0.03  $                               & $ 17.17   $      &        45$^{+16}_{-15}$    & $10^{+1}_{-1} $ &   $14^{+8}_{-6} $     \\ 
 including S*             & $ 0.08 \pm 0.03   $                              & $ 4.59   $         &          60$^{+26}_{-26}$   & $ 15^{+2}_{-3}$ &  $23^{+11}_{-9} $    \\ 
\hline
\end{tabular}}
\end{table}

\begin{figure}
\centering 
\includegraphics[width=7.5cm]{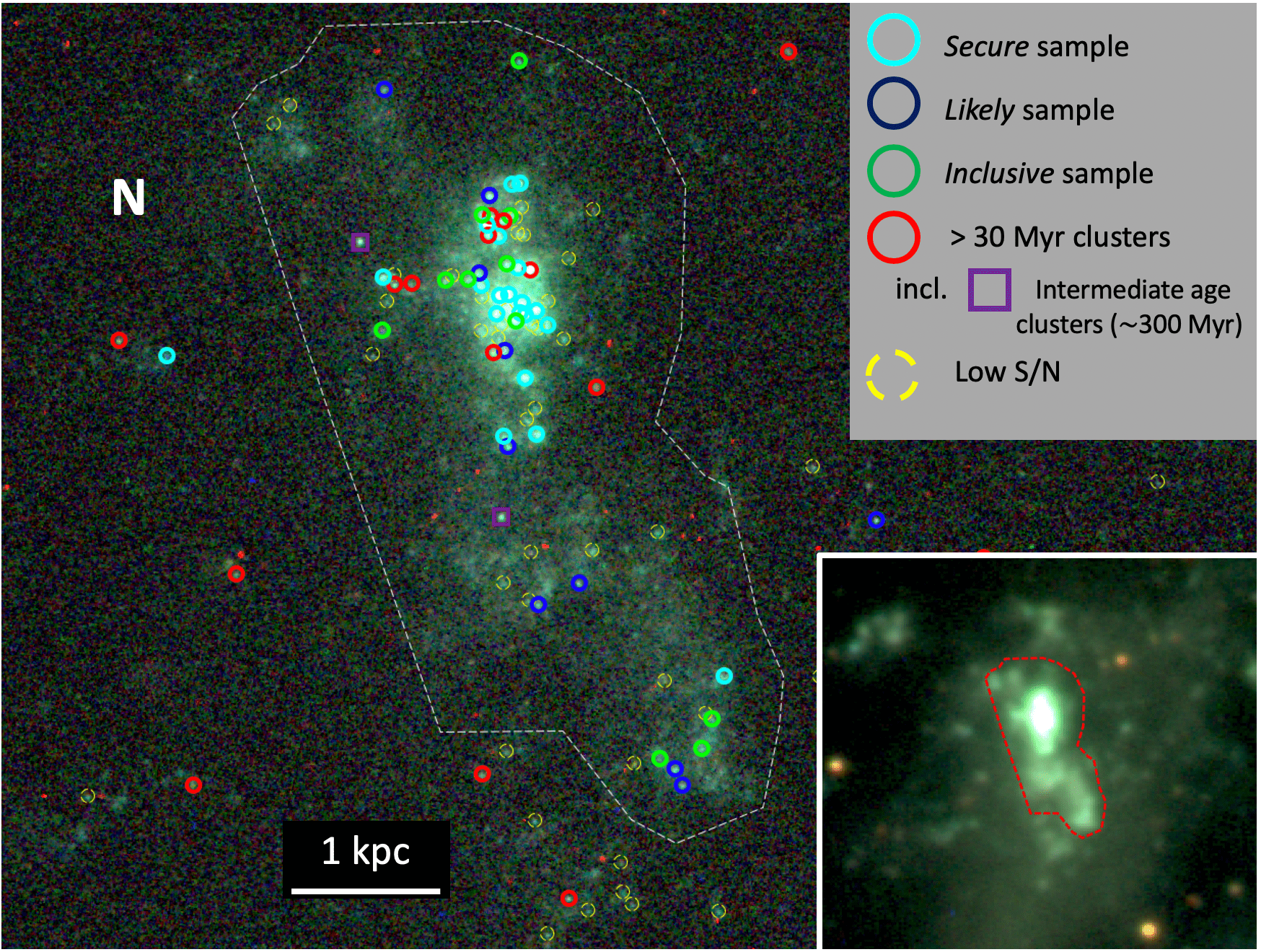}
\includegraphics[width=7.5cm]{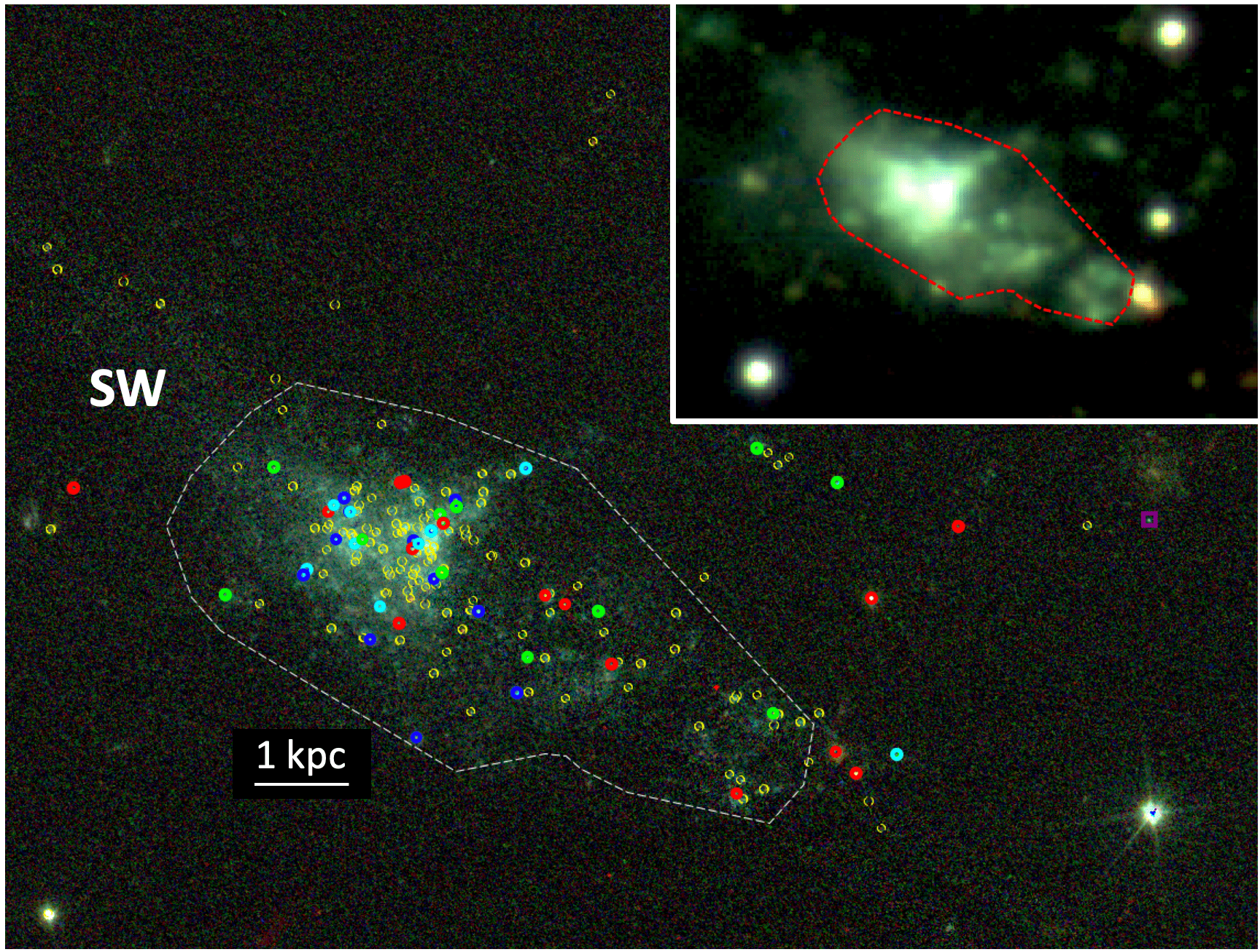}
\includegraphics[width=7.5cm]{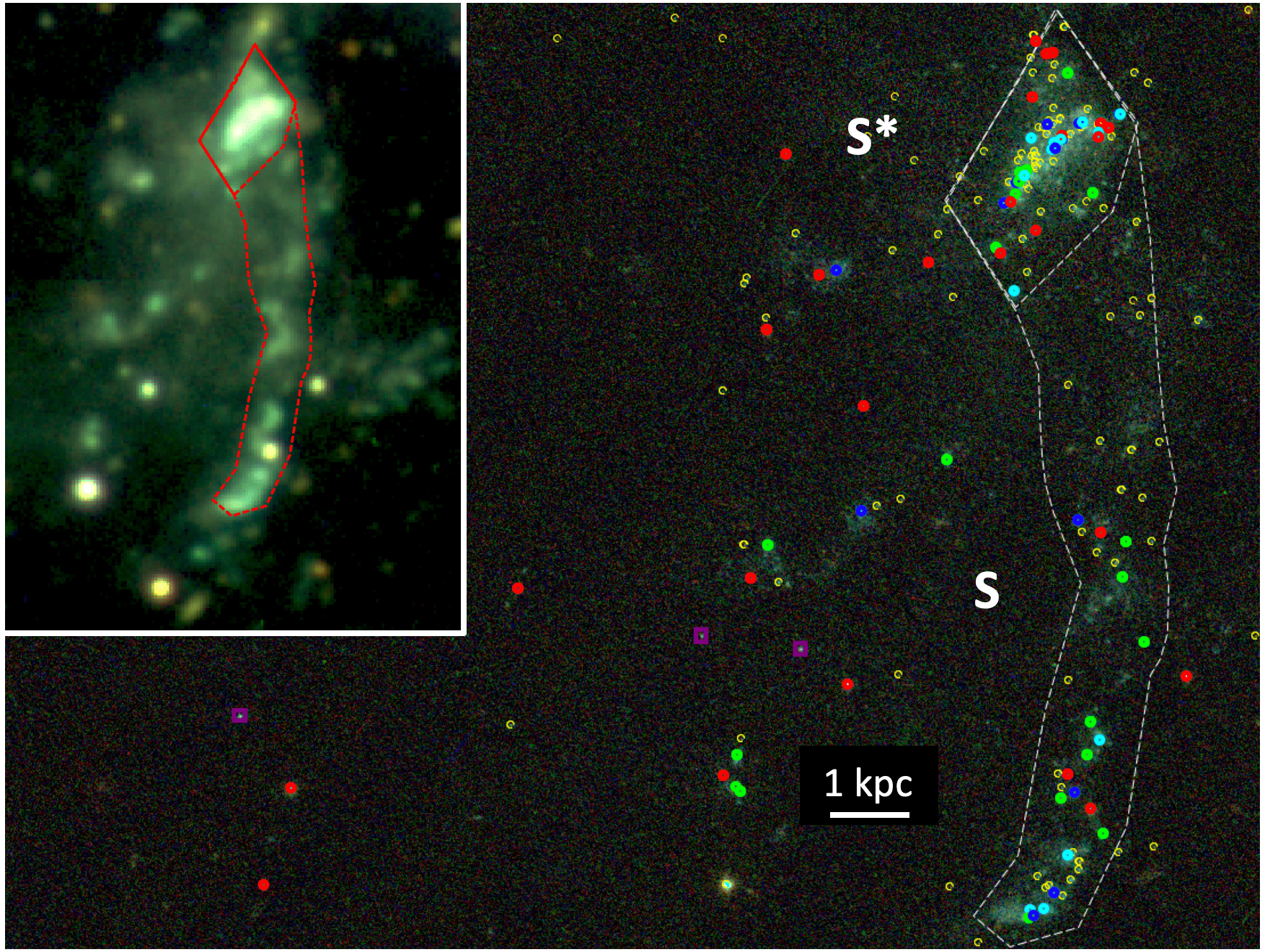}
\caption{True color image of the three TDGs: NGC~5291N (top panel), NGC~5291SW (middle panel) and NGC~5291S (bottom panel). The definitions of the young sample, degenerate clusters and older than 30~Myr samples are given in the text. The Intermediate clusters are part of the older than 30~Myr sample and are discussed in Sect.~\ref{res_int}. We also show detected clusters which do not have S/N > 3 in at least four bands, as the \emph{Low S/N} sample. The dashed white contours show the area considered to compute $\Sigma_\mathrm{SFR}$. The bottom panel show two contours: see text. The inset shows a VRI image from FORS \citep{Fensch16}, with the same contours. Only clusters inside the white contours are considered for the computation of the CFE. \label{map_gamma}}
\end{figure}

\begin{table}[]
\centering{
\caption{Significance, in standard deviations, of the offset of the data points compared to the three relations: \citet{Kruijssen12}, \citet{Johnson16} and \citet{Chandar17}, named respectively K12, J16 and C17.} 
\label{sigma_fid}
\begin{tabular}{l c c c }
\hline \hline
Galaxy               &   K12 & J16 & C17  \\  \hline 
TDG N               &   1.6   &  1.1 & 1.0   \\ 
TDG SW            &   1.3   &  1.0  & 0.5  \\
TDG S 		 &    2.2  & 1.7   & 1.1  \\
 including S*       &   1.7  &  1.3   & 1.3  \\ 
 TDG N+SW+S  &  3.8  &  3.1    & 2.5  \\
 TDG N+SW+S* &  3.5   &  2.8    & 2.6  \\
\hline
\end{tabular}}
\end{table}

One may characterise the \emph{cluster formation efficiency} (CFE) of galaxies by $\frac{\mathrm{CFR}}{\mathrm{SFR}}$, where CFR is the cluster formation rate (in $\Msun$/yr). It has been argued that galaxies follow a power-law relation with positive index in the CFE and the SFR surface density ($\Sigma_\mathrm{SFR}$) plane \citep{Larsen00, Billett02, Goddard10}. A similar relation was derived from theoretical grounds by \citet{Kruijssen12}. However, \citet{Chandar17} claim that the former empirical relation was driven by an under-estimation of the CFR of both the LMC and the SMC due to an inconsistent age range selection. On the contrary, they find a constant value of the CFE, of $24\% \pm 9\%$, independent of $\Sigma_\mathrm{SFR}$. \\ 

In order to compute the CFE for our system, we construct our young cluster samples based on the same definition as above, but for each of the three TDGs. In order to limit the effects of degeneracies, we use as a minimum value for the fitting prior A$_V$  > 0.3~mag, justified by the extinction maps obtained with MUSE by \citet{Fensch16}. We show the location of our young cluster sample in the three TDGs in Fig.~\ref{map_gamma}. We also show the detections that are \emph{degenerate} and \emph{securely old} (see definition in Sect.~\ref{young}). We also consider a smaller star cluster sub-sample for the S dwarf, shown in Fig.~\ref{map_gamma}, which we will call S* in the following. This is motivated by the elongated shape of this TDG, which suggests that TDG~5291S could actually be composed of two distinct objects, despite the apparent coherent HI rotation \citep{Lelli15}. \\

\begin{figure}
\includegraphics[width=9cm]{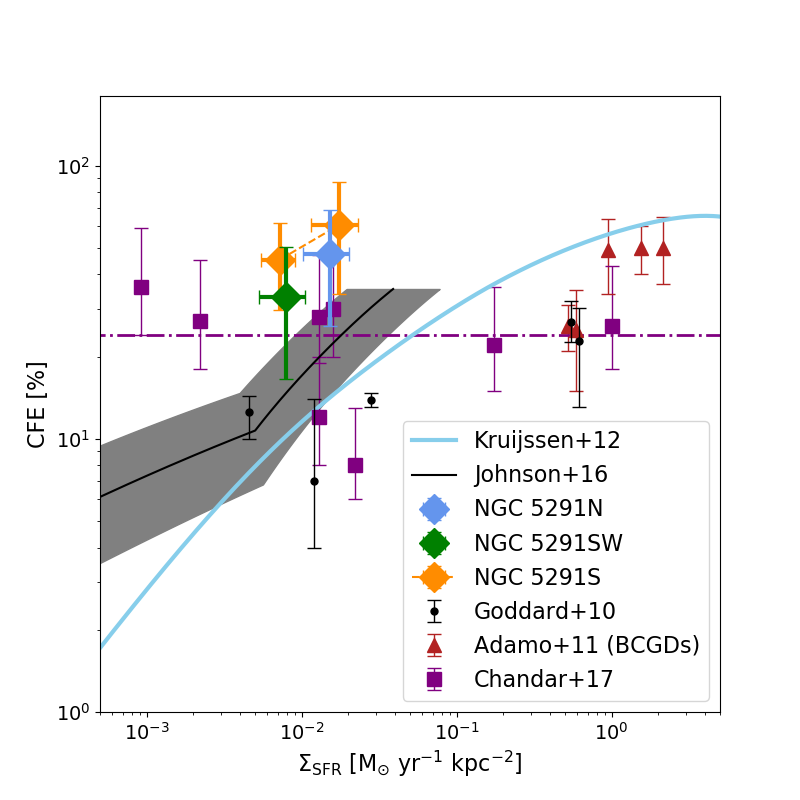}
\caption{Distribution of our TDGs in the CFE-$\Sigma_\mathrm{SFR}$ plane. For the TDG NGC~5291S we show two points, the full TDG (with lower CFE and $\Sigma_\mathrm{SFR}$) and only S*. The dataset of \citet{Goddard10} is shown in black. The sample of \citet{Chandar17}, and their fit to their data is shown in purple. For the SMC and the LMC we only show the values computed by the latter reference (see text). In red are shown the BCDGs of \citet{Adamo11}. The continuous blue line shows the prediction of the model by \citet{Kruijssen12} (see text).  The grey band shows a modified version of this model using the \citet{Bigiel08} relation \citep{Johnson16}.  \label{gamma}}
\end{figure}

Following previous studies on the CFE \citep[see e.g.][]{Goddard10, Adamo11}, we use the CMF to infer the total mass in clusters down to $10^{2}~\Msun$ by forcing a canonical power-law shape $\frac{dN}{dM} \propto M^{\alpha}$, with $\alpha = -2$ fit to the histogram. 

The SFR is obtained from the H$\alpha$ \citep{Boquien07}. 
This fixed index of -2 is also used in the studies we are referring to in this section, and is also supported by the shape of the CMF covering the full cluster sample (see Fig.~\ref{CMF}). Note that \citet{Boquien07} used a \citet{Salpeter55} IMF, whereas our mass estimates were obtained using a \citet{Chabrier03} IMF. We therefore multiply the SFR obtained using the H$\alpha$ by \citet{Boquien07} by a factor 0.70 to account for the different SFR to  H$\alpha$ flux ratio obtained for the two IMFs  \citep[see e.g.][]{Kennicutt09}. We also correct the SFR by the mean extinction measured in TDG~N, A$_V$ = 0.6 mag \citep[][]{Fensch16}. The obtained values of the CFE are summarized in Table~\ref{cfe_fid}. Using the lower metallicity prior (Z=0.004) give consistent results within the 1-sigma uncertainty and changes the CFE values by less than $10\%$. Using the 1~Myr star formation timescale changes the CFE values by less than 3\%.\\

To compare the TDGs with other star cluster forming galaxies, we place these values in the CFE-$\Sigma_\mathrm{SFR}$ plane in Fig.~\ref{gamma}. We see that the TDGs are located in the same regime as the BCDGs, with CFE above 45$\%$ for TDG N and TDG S. They are located systemically above the empirical \citet{Chandar17} relation although consistent within 0.5 to 1.3 $\sigma$.

In Figure~\ref{gamma} we show the current model and empirical predictions. The blue curve shows the model\footnote{Model accessible at: https://wwwmpa.mpa-garching.mpg.de/cfe/. We used the integrated CFE model.} by \citet[][K12 in the following]{Kruijssen12}, for a gas velocity dispersion of 30~km~s$^{-1}$. We also show the version of the model calibrated with the \citet{Bigiel08} relation between the SFR and the gas density \citep[][J16]{Johnson16}. In purple is shown the universal value of 24\% suggested by \citet[][C17]{Chandar17}. 
The computed CFE are systemically above these three relations. The significance of this deviation for the TDGs and the full system is measured with random draws of relation and data values, assuming gaussian distributions. For the combined TDGs, it is equivalent to multiplying the probabilities that each TDG's CFE has to be compatible with the relation. The significances are given in Table~\ref{sigma_fid}. While the measurement of each TDGs is less than 2.1$\sigma$ off of each relation, the combination of the TDG N, SW and S is above these relations by 3.8$\sigma$ for K12, 3$\sigma$ for J16 and 2.5$\sigma$ for C17. These numbers slightly change if one uses S* instead of S in the sample. Our sample of TDGs is then significantly above the current model and empirical relations.

We use an age range [1-30]~Myr which is broader than that typically used in these studies ([1-10]~Myr). We chose this range because there were not enough clusters younger than 10~Myr to properly measure the CFE. We did not correct for the mass evolution or destruction that may have happened, in particular cluster disruption by gas removal \citep[\emph{infant mortality,}][]{Boutloukos03, Whitmore07} which has a time-scale of 10-40~Myr \citep{Kroupa02, Fall05, Goodwin06}. This means that we are missing clusters which have been disrupted and mass which has been lost from the detected clusters. The fact that we do not correct for this effect suggests that we might be under-estimating the CFE of our TDGs (see discussion in C17). Finally we note that, at the distance of NGC~5291, we are contaminated by young star associations that are unbound and did not have time to dissolve \citep[see e.g.][]{Messa18}. This unresolved process might lead to an overestimation of the computed CFE. \\

As explained in Section~\ref{sub_deg}, we also performed the same analysis on two different samples of clusters with more restrictive or more relaxed age constraints. The analysis is presented in Appendix \ref{Annex_deg}. In particular, we introduce the {\it Secure} sample, which only contains clusters that are almost not affected by degeneracies and have a narrow age PDF, thus underestimating the genuine sample of clusters younger than 30~Myr. For this sample, the sample of TDGs (N,SW,S) is above the relations by 2.8$\sigma$ for K12, 1.8$\sigma$ for J16 and 1$\sigma$ for C17.
The fact that the CFE of the full sample is above the model relations of K12 by 2.8$\sigma$ confirms that this mismatch is robust against the age selection procedure. However, the CFE of the full sample of TDGs is only 1.8$\sigma$ from the J16 relation and is consistent with the C17 relation within 1$\sigma$. The combination of the CFEs of the TDGs are thus not statistically significantly above these two relations if one considers only this restricted sample.

Finally, we combined bands with different PSF. We added the F160W as it provides a good filter combination to reduce degeneracies \citep{Anders04}. However, the coarser spatial resolution of the F160W might lead to an over-estimation of the photometry in this band \citep[see e.g.][, but with aperture photometry]{Bastian14}. This effect will be discussed in section \ref{noIR}.


\subsection{Brightest cluster - SFR relation}
\label{res_Mv}

\begin{figure}
\includegraphics[width=9.5cm]{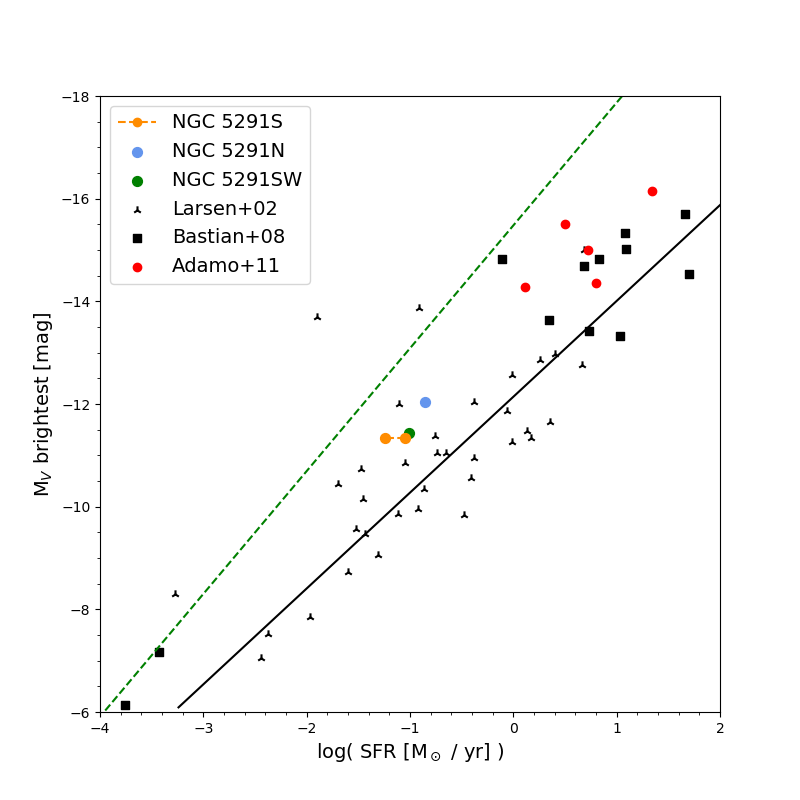}
\caption{Brightest cluster M$_V$-SFR relation for different galaxy samples. The NGC~5291S data point with the lowest SFR shows the value for S* only. The black line shows the fit of the \citet{Larsen02} sample. The dashed green line shows the maximum M$_V$ expected for a given SFR if all the star formation occurs in clusters with a $\frac{dN}{dM} \propto M^{-2}$ power-law \citep{Bastian08}. \label{Mv}}
\end{figure}

\citet{Larsen02} found a positive correlation between the V-band absolute magnitude ( $M_\mathrm{V}$) of the brightest cluster versus the SFR of the host, which was interpreted as a size-of-sample effect: the higher the SFR, the more clusters and thus the more likely high mass clusters would be found. The location of the three TDGs and other cluster forming systems in the $M_\mathrm{V}$-SFR plane is shown in Fig.~\ref{Mv}. We see that our three TDGs are located within the intrinsic scatter of the relation of \citet{Larsen02}. This suggests that the magnitude of the brightest cluster is a good tracer of the SFR for these systems, similar to what has been generally observed for star forming galaxies.

\subsection{Presence of intermediate age clusters}
\label{res_int}

It is interesting to study whether the peculiar environment of NGC 5291 may allow for the survival of clusters over timescales of $\simeq$100~Myr. \citet{Bournaud07} estimated that the interaction which triggered the formation of the ring happened around 360~Myr ago. As we have seen in Sect.~\ref{sub_deg} and Fig.~\ref{mass_age}, one may not estimate an age with a high precision. In Fig.~\ref{mass_age}, we can see that some cluster candidates with an estimated age between 100 and 2000~Myr have a relatively low r$_{\mathrm{age}}$ for the inferred age, that is, below 30. A large PDF is expected around these ages, as the stellar spectrum does not change much in this part of the stellar evolution period. Some of these clusters might therefore have formed at the time of formation of the ring and survived for several 100~Myr in this environment.\\

To construct a conservative sample of candidates with intermediate ages we first select clusters with P[50 < age < 2000] > 0.9. We chose an upper limit of 2000~Myr because the age PDFs can be quite extended for this age range (see Fig.~\ref{mass_age}). We note that this selection does not change if we allow for an extended star formation history with an exponential decrease timescale of 1 or 5~Myr, compared with our fiducial value of 0.1~Myr, chosen to model a quasi-instantaneous burst.\\

\begin{figure}
\includegraphics[width=9.5cm]{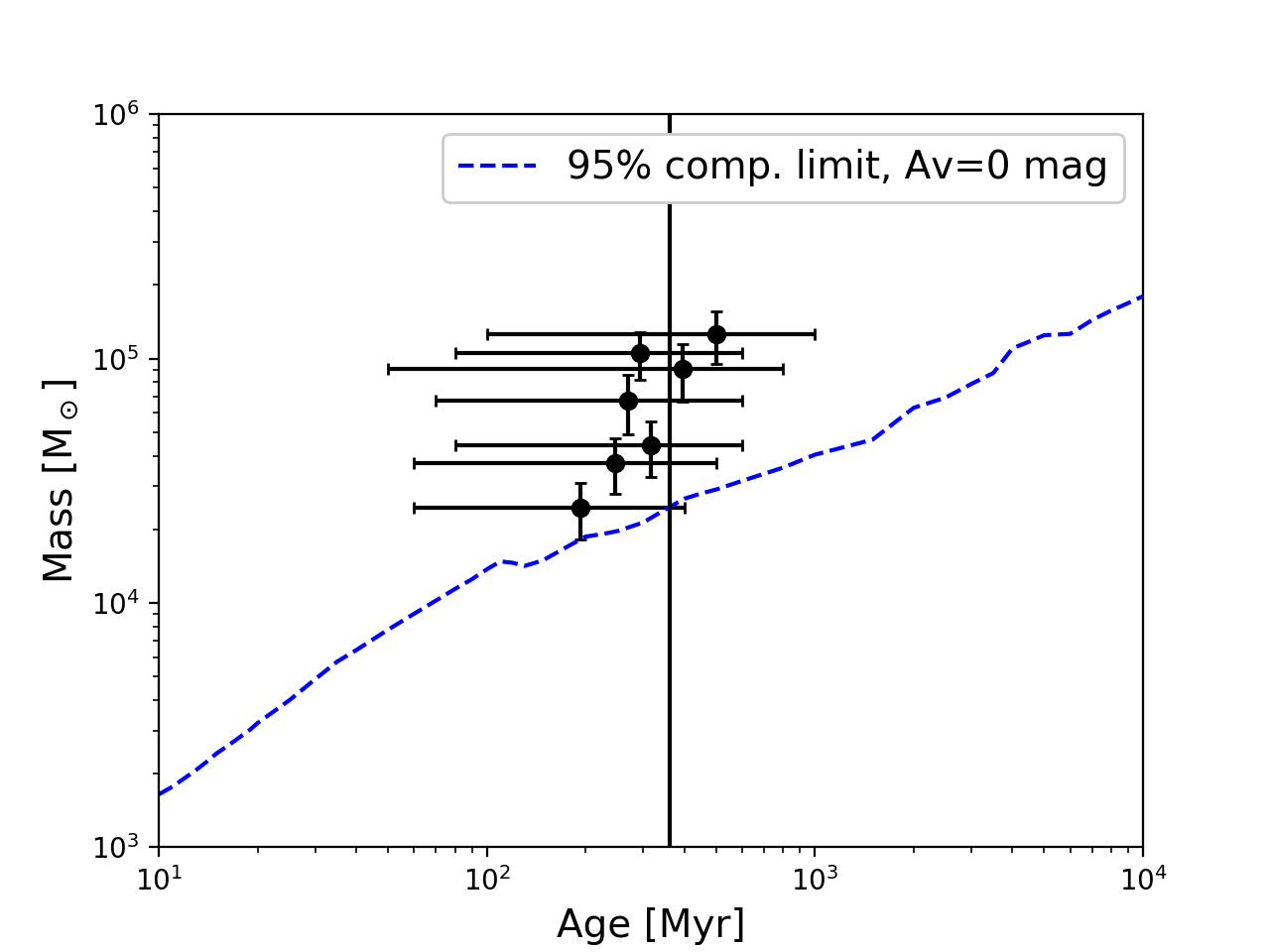}
\caption{Masses and ages of our conservative sample of intermediate age star clusters. The x-axis error-bar shows the width between the first and last decile of the age PDF. The y-axis error-bar shows the standard deviation for the mass estimate. The vertical black line shows the time of the formation of the ring structure, $\sim$360~Myr, as determined by \citet{Bournaud07}.  \label{int}}
\end{figure}

Moreover, we ensure that the photometry of these clusters is not consistent with them being old metal-poor or metal-rich GCs from the GC system of NGC~5291. For this, we run CIGALE with a broader metallicity prior (Z can be 0.0004, 0.004 or 0.008, instead of only 0.008), and we rule out clusters for which P[ age > 2500  ] > 0.1. We end up with seven clusters. Their ages and masses are shown in Fig.~\ref{int}. Their mass range is between $2\times10^{4}$ and $2\times10^{5} \Msun$. Their location is shown with purple squares in Fig.\ref{map_gamma}. Three are located close to the TDG N, one close to TDG SW and three close to TDG S. \\

One should note that clusters with similar masses and ages have already been seen in a number of dwarf galaxies \citep[see e.g.][]{Larsen04, deGrijs13}. However, their presence in NGC~5291 shows that massive star clusters can survive the very turbulent and gaseous environment of a tidal dwarf galaxy from its formation up to hundreds of millions of years.

\section{Discussion}
\label{Discussion}

\subsection{Effect of including the F160W band}
\label{noIR}

\begin{table}[]
\centering
\caption{Second column shows the values of the CFE for the three TDGs, without the F160W band. The three last columns show the significance, in standard deviations, of the offset of the data with respect to the models.  \label{table::CFE_noIR}}
\label{my-label}
\begin{tabular}{l c c c c c} \hline \hline
Galaxy      &  CFE [$\%$]  & K12 & J16 & C17\\  \hline
TDG N       &  37$^{+15}_{-15}$  & 1.5 & 0.8   &  0.7 \\ 
TDG SW    & 26$^{+11}_{-10}$   & 1.4 & 0.8   &   0.1 \\ 
TDG S       &   33$^{+14}_{-14}$ & 1.6 & 1.2 &  0.5 \\
including S* & 46$^{+25}_{-25}$  & 1.2 & 0.8  & 0.8 \\
TDG N+SW+S  &  -  & 3.5 & 2.6 & 1.9 \\
TDG N+SW+S* &   -  & 3.3 & 2.4 & 2.0 \\ \hline 
\end{tabular}
\end{table}

\begin{figure}
\includegraphics[width=9cm]{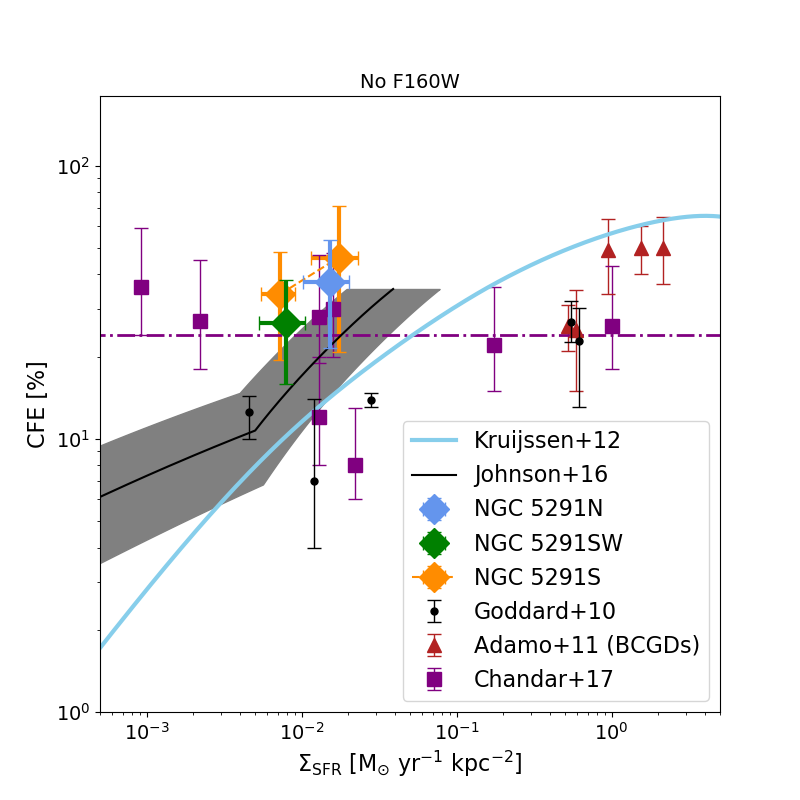}
\caption{Same legend as Fig.~\ref{gamma}. The analysis was done without considering the F160W band. \label{CFE_noIR}}
\end{figure}

\begin{figure}
\includegraphics[width=9cm]{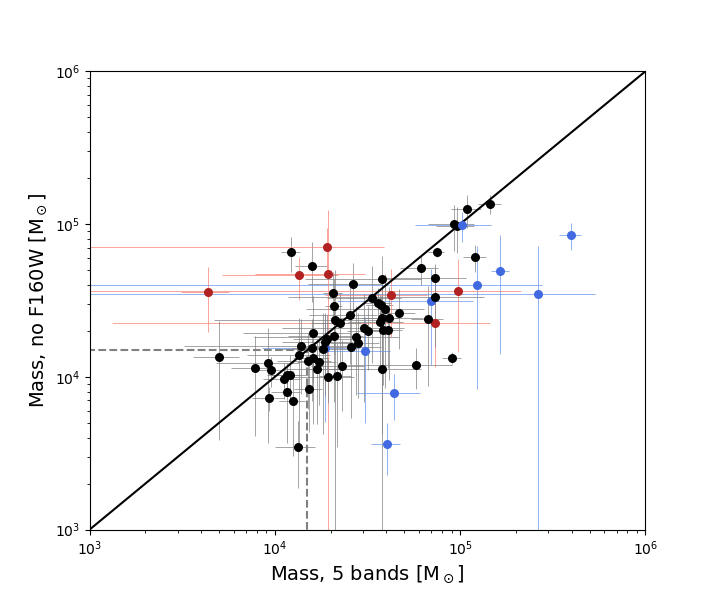}
\caption{Comparison between the mass estimation between the analysis including or not the F160W band. The blue points show clusters that are included as {\it younger than 30~Myrs} only in the analysis without F160W. The red points show clusters that are included in this category only when the F160W band is included. The thick line shows the identity function. The two dashed lines show the mass completeness limit. \label{mass_change}}
\end{figure}

In Section 4.4 we saw that two TDGs, N and S, have very high CFEs, with an average value of 42 \%. While it has been argued on theoretical grounds that star cluster formation should be more efficient in low-metallicity environments, all other factors being equal \citep{Peebles84, Kimm16}, these metal-rich TDGs reach a similar CFE of metal-poor BCDGs.

As pointed out in Section~\ref{res_gamma}, the estimated masses could have been affected by an overestimation of the flux in the F160W band. We included the F160W filter to reduce the degeneracies on the estimation of the age and mass of the clusters \citep{Anders04}. However, this band has a coarser spatial resolution than the four other ones. The F160W band PSF has a FWHM of about 0.18\arcsec, while the four other bands have PSF FWHMs of 0.06\arcsec. Some F160W flux measurements might have been contaminated by regions that are very close to the clusters and that are not included in the other bands, which might affect the derivation of the physical quantities \citep[e.g.][in the case of aperture photometry]{Bastian14}.

To test the effect of adding the F160W band, we removed it from the analysis for the measurement of the CFE. The CFEs we obtain are summarised in Table~\ref{table::CFE_noIR}. They are lower than the ones obtained with the five bands, by typically 22\%. 

In Fig.~\ref{mass_change} we see the change in the mass estimation of the clusters that are considered as {\it younger than 30 Myrs} and located in the TDGs. Of the clusters we that determined to be younger than 30 Myr, a few are estimated to be older than this limit when ignoring the F160W filter (5 of them above the completeness limit) while others are estimated to be younger with the F160W band (8 above the limit). Second we see that, while the estimated masses are very similar, there is a trend towards lower masses if one does not consider the F160W band. 

In Fig.~\ref{CFE_noIR} we see that even without the F160W band the CFEs of the TDGs are systemically above the three relations from the literature. The computation of the significance of the discrepancies are given in Table~\ref{table::CFE_noIR}. The full sample of TDGs (N,SW,S) is respectively 3.5$\sigma$ and 2.6$\sigma$ above the K12 and J16 relations, that is with more than 99.5\% certainty. However, the offset with the C17 relation is only 1.9$\sigma$.
Thus, the exclusion of the F160W reduces the discrepancy between the data and the relations from the literature, which becomes statistically insignificant only for C17 (< 2.5$\sigma$). This suggests that the discrepancy to the three relations from the literature is robust against over the possible overestimation of the F160W band due to a coarser spatial resolution.

\subsection{What is the origin of the high cluster formation efficiency ?}
\label{disc_gamma}

In Section~\ref{res_gamma} we saw that our sample of TDGs have high CFEs, above 30\%, similar to what is observed for BCDGs While it has been argued on theoretical grounds that star cluster formation should be more efficient in low-metallicity environments, all other factors being equal \citep{Peebles84, Kimm16}, these metal-rich TDGs reach a similar CFE of metal-poor BCDGs. \\

It is interesting to note that all galaxies from the literature which have a CFE above $20\% $ are galaxies involved in an interaction of some sort, with the exception of the center of M~83 in the sample of \citet{Goddard10}. It is also interesting to note that the late stages of mergers, while leading to similar $\Sigma_\mathrm{SFR}$ as the early stages, triggers the formation of only a few clusters \citep{Renaud15}, and have thus a much lower CFE. Their interpretation is that cluster formation is triggered by the onset of compressive turbulence which is triggered mainly during the early times of galaxy interaction.\\

Although the TDGs in NGC~5291 were not formed in bona fide merging galaxies, they are located in a gas-dominated environment and are probably not fully relaxed yet. It is thus possible that their dynamical state, in terms of compressive turbulence, is similar to that of interacting galaxies, possibly because of accretion from the gas ring \citep{Fensch16}.\\

\subsection{Evolution of the star cluster system}
\label{disc_int}

We saw in Section~\ref{res_int} that some clusters could survive their birth environment for several $100$~Myr. The fact that we could find some in the gaseous ring shows that we can expect the survival of massive star clusters from the formation of the tidal dwarf galaxy to at least several $100$~Myr.\\

One may now wonder how this specific star cluster system will evolve in the future. In the following, we consider that our clusters survived after gas expulsion and we do not consider their \emph{infant mortality} rate, which is due to the internal feedback expelling the gas and destabilizing the cluster and which has a timescale of 10-40~Myr (see Sect.\ref{young}). We model the mass loss due to cluster evaporation during its relaxation as $\Delta$M(t) = $\mu_{\mathrm{ev}} t$ with $\mu_{\mathrm{ev}}$ the evaporation rate \citep{Henon61, Fall01, Jordan07}, given by:

\begin{equation}
\mu_{\mathrm{ev}} = 345~\Msun~\mathrm{Gyr}^{-1} \Big( \frac{\rho}{\Msun~\mathrm{pc}^{-3}}\Big) ^{1/2}
\end{equation}

where $\rho$ = 3M / (8$\pi$R$_{\mathrm{eff}}^{3}$) is the half-mass density of the cluster, with M and R$_{\mathrm{eff}}$ its mass and half-mass radius. This is a likely lower limit to the genuine evaporation rate of stellar clusters, as it does not include the effects of stellar evolution, gas cloud encounters and tidal effects from the host TDG.\\ 

The typical density of YMCs is $10^3~\Msun$~pc$^{-3}$ \citep[see review by][]{Portegies10}. For such a density, we obtain $\mu_{\mathrm{ev}} \sim 10^4~\Msun~\mathrm{Gyr}^{-1} $. Under this hypothesis, one may conclude that most of the stellar clusters of our system will be destroyed in a few Gyr at most, at least by internal relaxation. \\

Now consider a cluster with a mass of $2\times10^4~\Msun$. Reaching the typical density of YMCs implies a typical half-mass radius of 1.3~pc. As one pixel corresponds to a physical size of 12~pc (36~pc for the F160W band) at the distance of NGC~5291, we cannot constrain the size of our clusters. Most of our sources are well fitted by a PSF, which means a half-mass radius securely below 6~pc. The few sources which are not well fitted by a PSF (see Sect.~\ref{Technics}) have half-light radii which can reach up to 2.5~pixels. However, they could also be blended detections or extended nebular emission from the ionized outskirts of young clusters. For a mass of  $2\times10^4~\Msun$ and a half-mass radius of 6~pc, we obtain $\mu_{\mathrm{ev}} \sim 10^3 ~\Msun~\mathrm{Gyr}^{-1}$. The timescale for the destruction -- considering only the effects of relaxation -- is therefore a Hubble time for $2\times10^4~\Msun$ clusters. Under this hypothesis it is possible that the most massive clusters of in sample (reaching typically $2\times10^5~\Msun$) can survive evaporation from internal processes for several Gyr. However, the disruption from tidal effects will be faster \citep{Gieles06}.\\

Mass loss from stellar evolution (40\% to 60\% of the mass over 12~Gyr; \citealt{Kruijssen08, Sippel12}), from gas cloud encounters and from tidal harassment still needs to be taken into account. The study of these two mass loss processes goes beyond the scope of the present study. Given the very gaseous environment of these TDGs and the high gas turbulence, one may expect the latter two processes to be more efficient than in an isolated and kinematically relaxed systems. At the same time, star clusters in such a gas-rich environment may continue to accrete gas from their surroundings \citep{Pflamm09, Naiman11, Li16}.\\

Thus, if the YMCs of these TDGs are similar to YMCs observed in other environments in terms of density, we do not expect these dwarfs to form a system of massive star clusters which could last for a Hubble time. This conclusion is mitigated if we allow for a lower star density, which remains empirically unconstrained given our spatial resolution.

  
 \subsection{Evolution of the TDGs}
 
 The $\Lambda$CDM paradigm predicts a different dark matter (DM) content for two classes of dwarf galaxies: TDGs, formed during interactions and which should be devoid of DM, and \emph{normal} dwarf galaxies, such as the dwarf ellipticals (dE), dwarf irregulars or BCDGs, formed inside a DM halo. This difference in DM would result in different kinematics and provide us with a new test for the $\Lambda$CDM paradigm \citep[see e.g.][]{Kroupa10}. Even though the absence of dark matter in TDGs is predicted from numerical simulations \citep{Bournaud06,Wetzstein07, Bournaud08}, it is hard to prove as these are young and turbulent systems. Under the assumption of dynamical equilibrium, suggested by simulations to occur in less than one orbital time \citep{Bournaud06, Bournaud07}, HI kinematics are consistent with a purely baryonic content \citep{Lelli15}. One needs to investigate the kinematics of old TDGs, which are kinematically relaxed, to confirm such a purely baryonic content, which requires to distinguish TDGs from dEs.  

TDGs are known to be outliers of the luminosity-metallicity relation \citep[e.g.][]{Duc00, Weilbacher03}. However for old, gas-poor TDGs, obtaining the metallicity from the stellar population might still be very challenging with current observing facilities. Moreover, as the metallicity of the host is a decreasing function of redshift, one may argue that the deviation from the magnitude-metallicity relation will decrease, making it harder to separate old and equally aged TDGs and dEs. TDGs are also known to be outliers from the size-mass relation for dwarf galaxies, having unusually large effective radii for their mass \citep{Duc14}.\\

A final means to distinguish these two categories could be to use their stellar cluster content \citep{Dabringhausen13}. dEs are known to host a significant number of GCs compared to their mass, with specific frequencies reaching up to 100 \citep{Peng08, Georgiev10}. Our analysis showed that even quite massive star clusters may form in TDGs. Some may be able to survive for several Gyr and thus be visible in rather old TDGs, but they will likely evaporate within a Hubble time. This is due to the fact that their SFR is too low to form clusters that are massive enough to survive evaporation for several Gyr. For a Hubble time, the minimum mass would be around the turn-over value for the GC mass function (GCMF), $2\times 10^5~\Msun$ \citep{Fall01, Jordan07}. \\

Moreover, as a TDG potential well does not trap significant amounts of DM or old stars from the host, one may argue that the capture of GCs from the host, which are kinematically coupled to either the DM halo or the bulge component \citep[see review by][]{Brodie06} is also unlikely. This will be verified for our system in a future paper which will focus on the old cluster population, as described in Section~\ref{Obs}. The accretion of old GCs onto TDGs also needs to be investigated by means of numerical simulations to understand the effect of varying the orbital parameters. \\

However the conditions at higher redshift are most likely different, as the host galaxy is likely to have a more substantial gas component \citep[see e.g.][]{Combes13}. Among the rare literature on TDG formation at high redshift, simulations by \citet{Wetzstein07} showed that more gas-rich disk galaxies are more likely to form TDGs, and \citet{Elmegreen07} found five young TDG candidates at $z = 0.15-0.7$, which have higher stellar masses than typical local TDGs (up to $5\times10^9~\Msun$).  As claimed by the latter, the higher velocity dispersion of both the gaseous and stellar components of higher redshift galaxies could lead to Jeans masses of up to $10^{10}~\Msun$ in tidal tails. 

Thus, one may argue that at a given higher redshift TDGs will have higher gas masses and higher SFRs. If star cluster formation at this cosmic epoch follows the empirical relation between the SFR of a galaxy and the magnitude of its brightest star cluster, given that the stellar models we used predict a M$_\mathrm{V} = -12.4$~mag for a 10~Myr old cluster of $2\times10^5~\Msun$, then a SFR of 5-10$\Msun$ yr$^{-1}$ would be sufficient to form some clusters more massive than the peak of the GCMF, which would be able to survive cluster dissolution for a Hubble time. \\

Although our analysis shows that TDGs formed under the current conditions are not likely to keep a GC system, more investigation is needed to understand if TDGs formed at higher redshifts would be able to harbour a GC system until the present epoch, and if one could distinguish them from other dwarfs using this criterion. Recently, a UDG candidate, DF2 and DF4, were found sharing several of the properties expected for TDGs: a putative lack of DM, a large effective radius and the proximity of a massive galaxy \citep{vanDokkum18a}, which led to speculation of a tidal origin. Note that the DM of this galaxy is still the subject of intense debate in the community \citep[see e.g.][]{Martin18,Trujillo18, Blakeslee18, Emsellem19, Danieli19}. However one unique feature is its large number of massive GCs \citep{vanDokkum18b}. \citet{Fensch19}, found that the metallicity of the stellar body of DF2 and its GCs could be consistent with DF2 being an old TDG. However, a massive TDG like those around NGC~5291 did not form such massive clusters.

 \subsection{Link to the formation of GCs in high-redshift galaxies}
 
 A prevailing theory for the formation of the metal-rich population of GC around present-day massive galaxies is that they may have formed in the star-forming disk of the host galaxy at high-redshift  \citep{Shapiro10, Kruijssen12}, when their morphology was dominated by 5 to 10 UV-bright giant clumps (mass $\sim 10^{7-9}~\Msun$, radius $\sim~1-3$~kpc, \citealt{Cowie96, Elmegreen09}). A resolved study of clustered star formation in these clumps is unfortunately not possible with current instrumentation, except in some fortuitous cases of strong gravitational lensing \citep{Cava18}. Thus local analogues are often used as laboratories to investigate the possible ISM and star cluster formation, such as the nearby BCDGs \citep{Elmegreen12}. In particular, they have been shown to be very efficient at forming YMCs \citep[][and Section~\ref{res_gamma}]{Ostlin03, Adamo10, Lagos11}. Although  BCDGs are characterized by high gas fractions and turbulence \citep[see e.g.][]{Lelli14}, similar to what is expected for higher redshift galaxies. However they usually have low to very low metallicities (typically 0.2~Z$_\odot$, \citealt{Zhao13}), while the giant clumps at high redshift already reach moderate metallicity, between 1/3 and 1/2 solar \citep{Erb06, Cresci10, Zanella15}, which may have a strong impact on gas fragmentation \citep{Krumholz12}. Moreover, most of their stellar mass resides in an old stellar component \citep{Loose86, Papaderos96}, making these systems dynamically old. \\
 
TDGs are gas-rich, dynamically young and have moderate metallicity, and thus should be better analogues to the clumps of high-redshift galaxies. The high to very high CFEs (up to 50$\%$) observed in the TDGs presented in this study suggest that the physical conditions in high-redshift galaxies could be very favorable to the formation of star clusters. Moreover, if the empirical relation between the SFR of a galaxy and the magnitude of its brightest star cluster holds at these redshifts, since giant clumps have SFRs of about $1-10~\Msun$ yr$^{-1}$ \citep{Guo12}, one may expect them to produce star clusters more massive $2\times10^5~\Msun$, the likely threshold mass which would allow them to survive dissolution over a Hubble time (see previous subsection).\\

It should be noted that the molecular surface gas density of TDGs is much lower than that of high-redshift galaxies, by 2 orders of magnitude \citep{Lisenfeld16}, and their depletion timescale is higher by a factor of 10 (2~Gyr for TDGs, \citealt{Braine01}, 0.2~Gyr for $z \simeq 2$ galaxies, \citealt{Combes13}). Moreover, the tidal forces from the host are likely different, and are important for the formation of YMCs in colliding galaxies \citep{Renaud15}, as well as for their survival \citep{Baumgardt03, Renaud11}. Numerical simulation work is therefore still needed to understand cluster formation in the giant clumps of high-redshift galaxies.\\

\section{Conclusion}
\label{Conclusion}

We investigated star cluster formation and evolution in three tidal dwarf galaxies, whose physical properties differs from the ones of starbursting dwarfs. In particular, they are gas-rich, highly turbulent and have a gas metallicity already enriched to up to half-solar. \\

The three TDGs are located in a huge collisional ring around NGC~5291. We observed this system with the \emph{HST} using five broad bands from the near-UV to the near-IR. The photometry was extracted using PSF and S\'ersic-fitting, and we compared the obtained SED with stellar evolution models using the CIGALE code. \\

{We find that star clusters are observed in TDGs, with masses of up to $10^5~\Msun$, with a mass distribution similar to those observed in other star cluster forming systems. After taking into account the effect of the extinction-age degeneracies, we studied the star cluster formation efficiency in the TDGs. We showed that the three TDGs have high CFEs, above 30\%, with an average of 42\%. This is comparable to BCDGs, but with a lower SFR surface density, a higher metallicity and without being bona fide merging systems. The full sample of TDGs is located 2.5 to 3.8 $\sigma$ above the relations from the literature. There may be uncertainties not yet recognised which still allow a constant CFE at this time \citep[see e.g.][]{Chandar17}, and more data is needed for similar special type of galaxies. Nevertheless, our results suggest that such a constant CFE relation would have a large scatter, and that there would be structure within this scatter, correlated with galaxy type and/or environment.

We next probed the existence of intermediate age clusters, which could have formed during the early stages of the formation of the gaseous ring structure and may have survived for several 100~Myr. The fact that we could find some of them shows that cluster formation started early and  we can expect the survival of young massive (above $10^4\Msun$) star clusters from the formation of their host dwarf to several $100$~Myr. However, if they have a similar density to what is observed for YMCs in other known environments (BCDGs, mergers), they might be present for a few Gyr but destroyed in a Hubble time because of relaxation-driven dissolution effects. If TDGs formed at high redshift have a higher SFR, we may expect them to form more massive clusters that would be able to survive cluster dissolution for a Hubble time.

\begin{acknowledgements}
The authors thank the referee for their very useful comments which helped improve the paper.
MB acknowledges support of FONDECYT regular grant 1170618.
Support for Program number HST-GO-14727 was provided by NASA through a
grant from the Space Telescope Science Institute, which is operated by the
Association of Universities for Research in Astronomy, Incorporated, under NASA
contract NAS5-26555. DME  acknowledges support from grant HST-GO-14727.002-A. B.G.E. acknowledges support from grant HST-GO-14727.004-A. EB acknowledges support from the UK Science and Technology Facilities Council [grant number ST/M001008/1]. FR acknowledges support from the Knut and Alice Wallenberg Foundation.
\end{acknowledgements}

\bibliographystyle{aa}
\bibliography{example}

\appendix

\section{Single band images for TDG SW and S}
\label{appendix::fits}

The images corresponding of Fig.~\ref{map} for TDG S and SW are presented respectively in Fig.~\ref{map_S} and Fig.~\ref{map_SW}.

\begin{figure*}
\centering
\includegraphics[width=5.0cm]{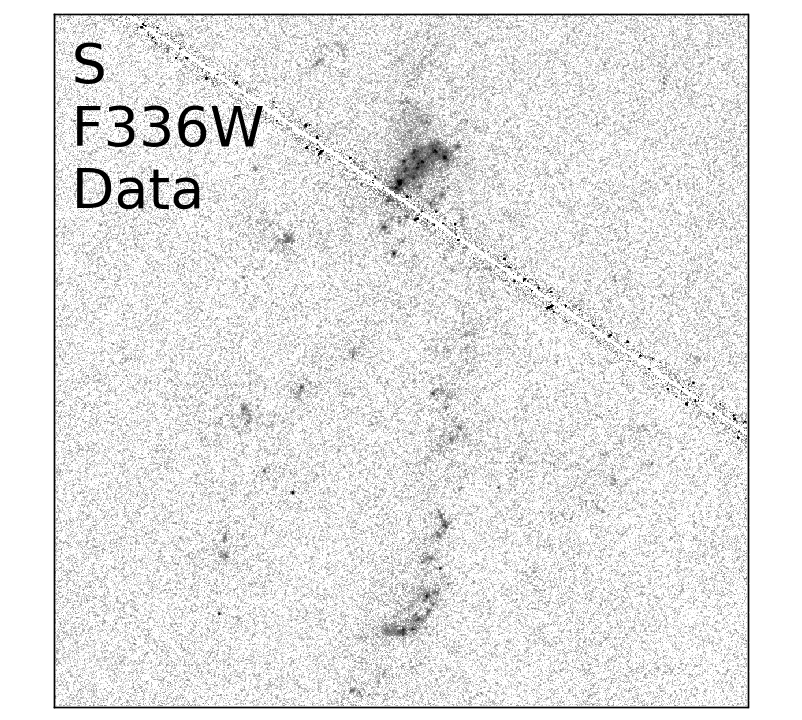}
\includegraphics[width=5.0cm]{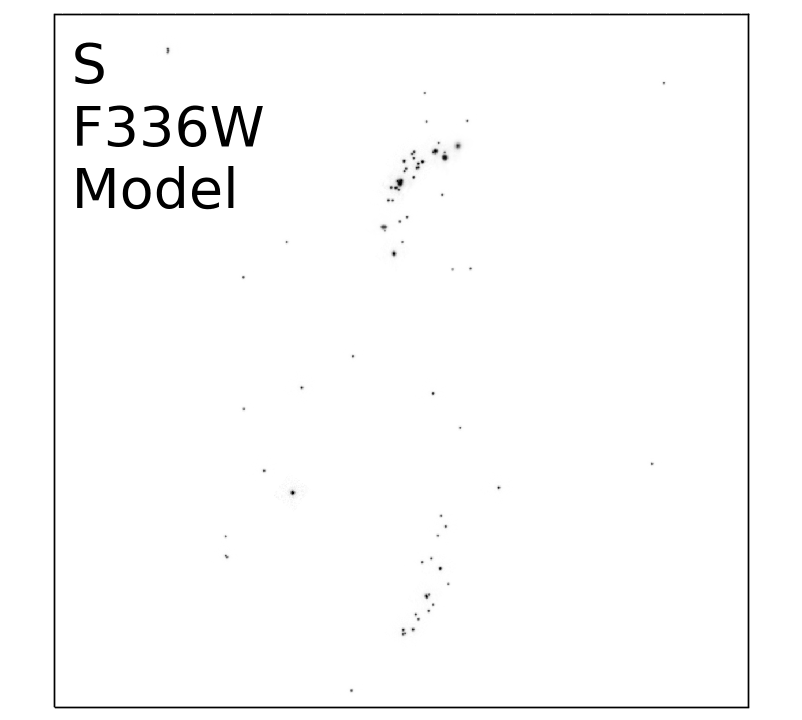}
\includegraphics[width=5.0cm]{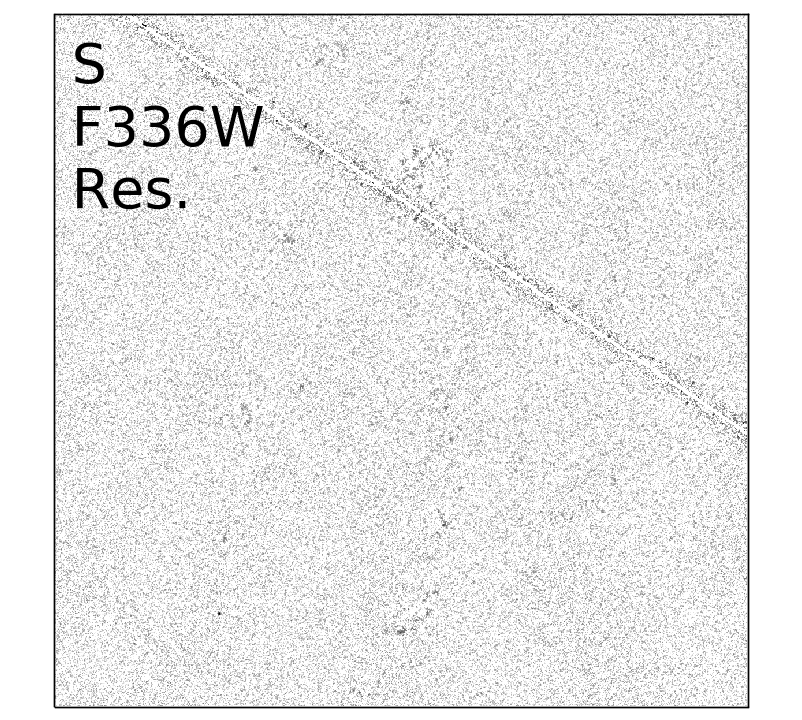}\\
\includegraphics[width=5.0cm]{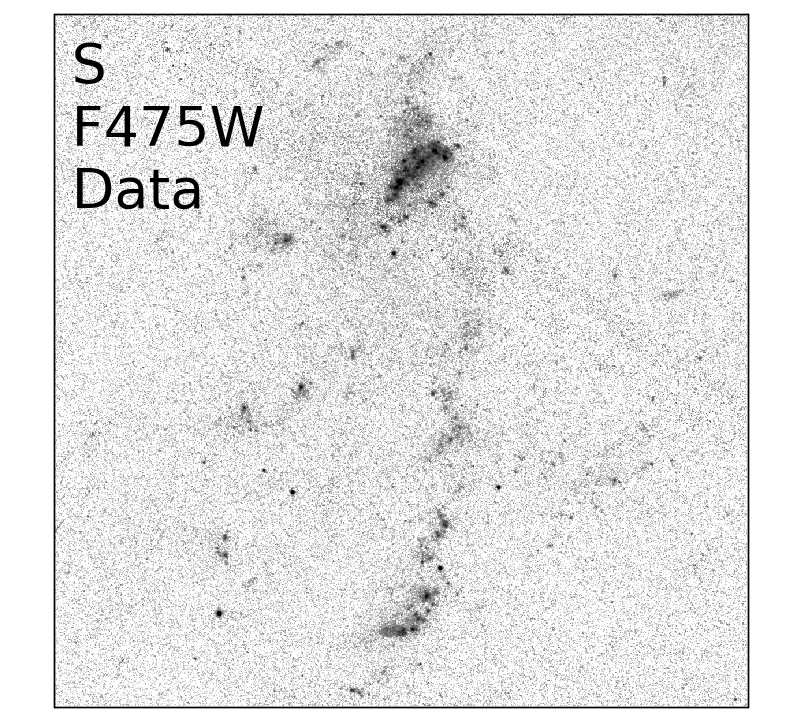}
\includegraphics[width=5.0cm]{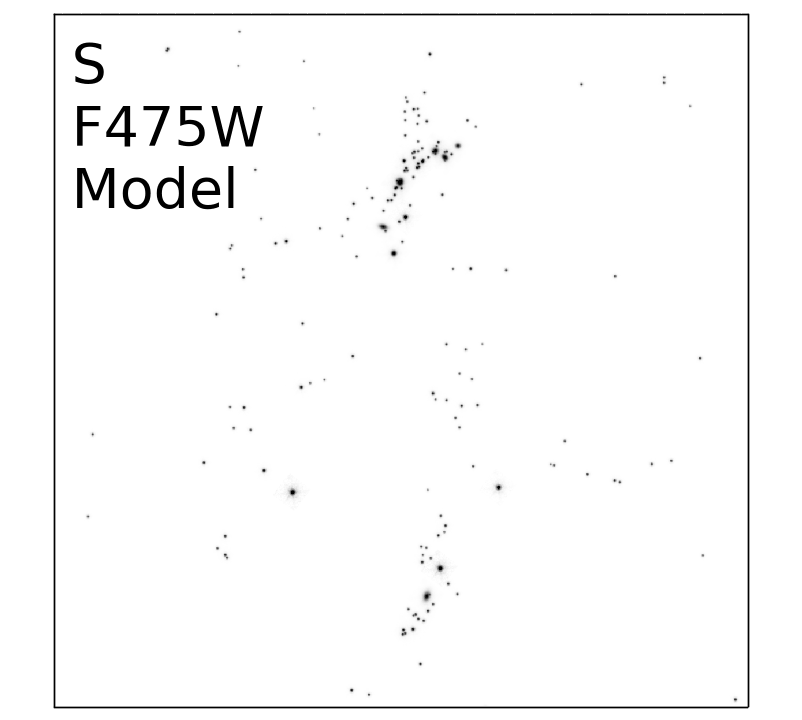}
\includegraphics[width=5.0cm]{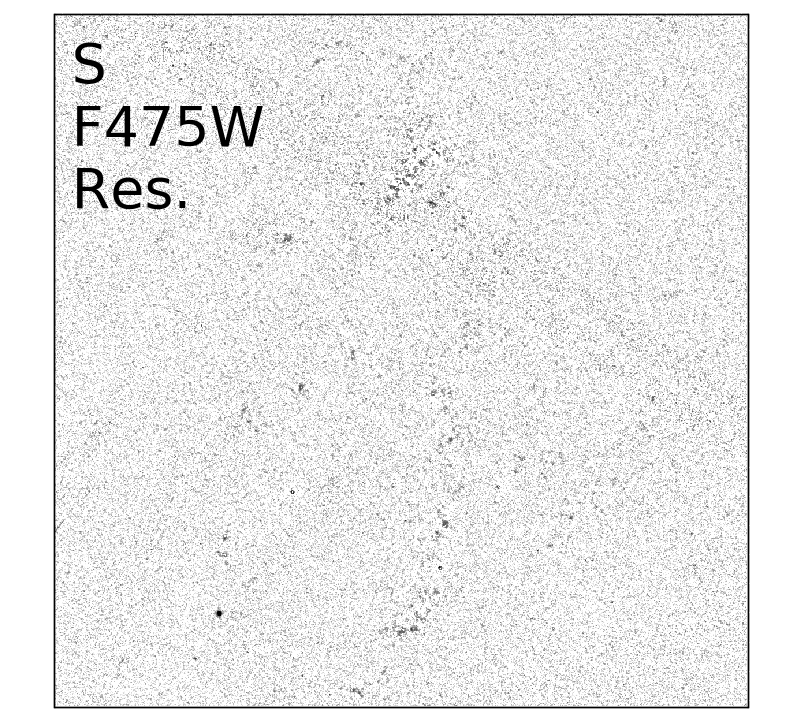}\\
\includegraphics[width=5.0cm]{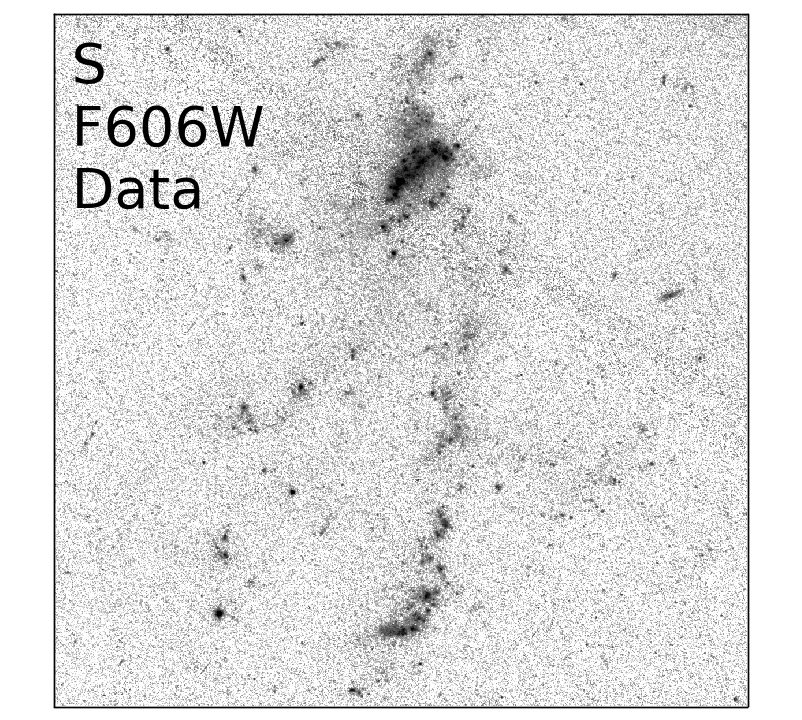}
\includegraphics[width=5.0cm]{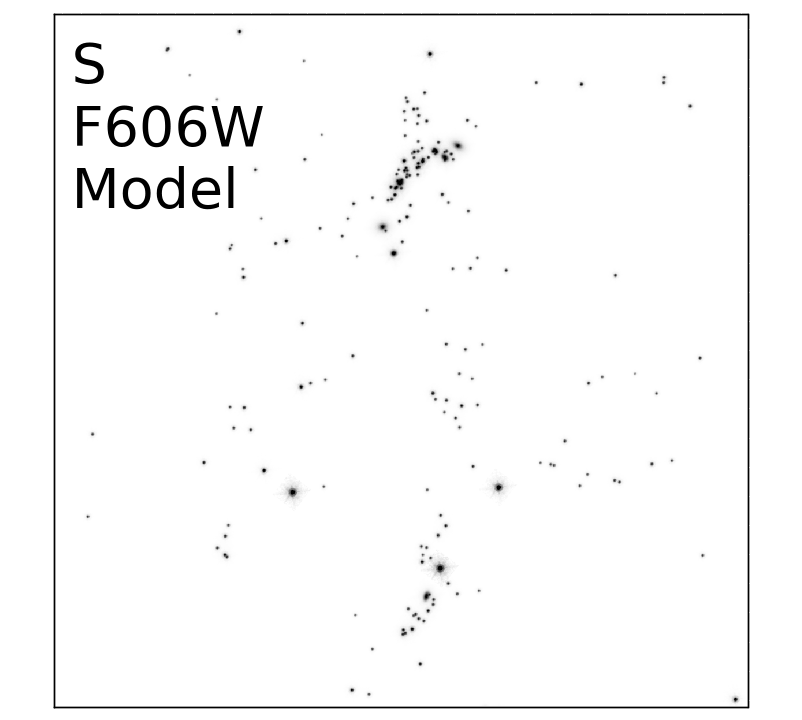}
\includegraphics[width=5.0cm]{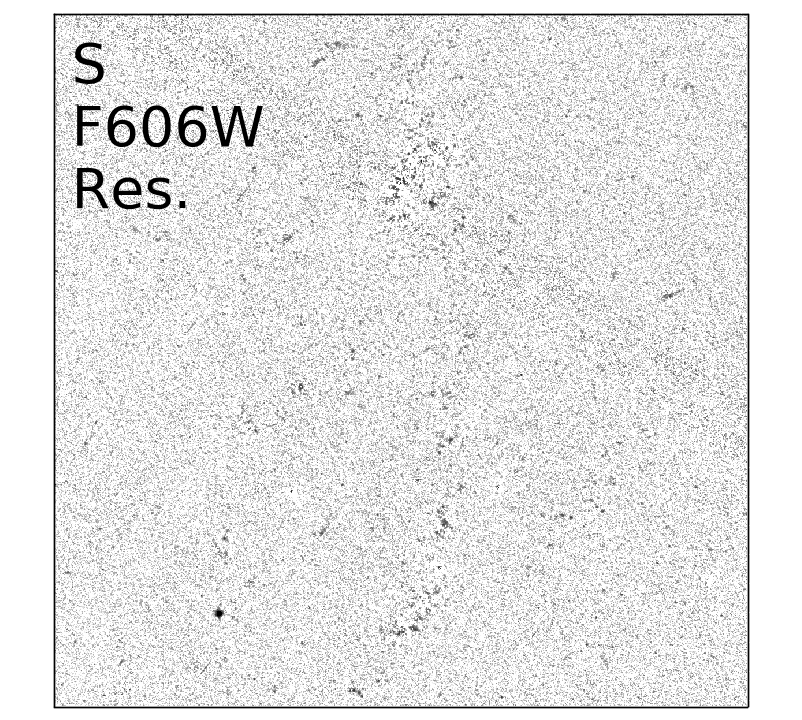}\\
\includegraphics[width=5.0cm]{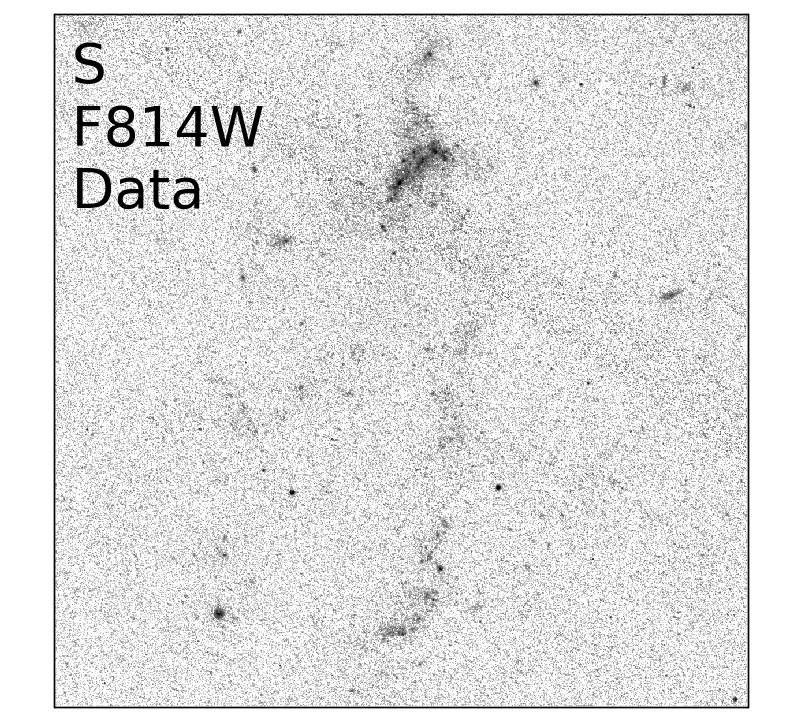}
\includegraphics[width=5.0cm]{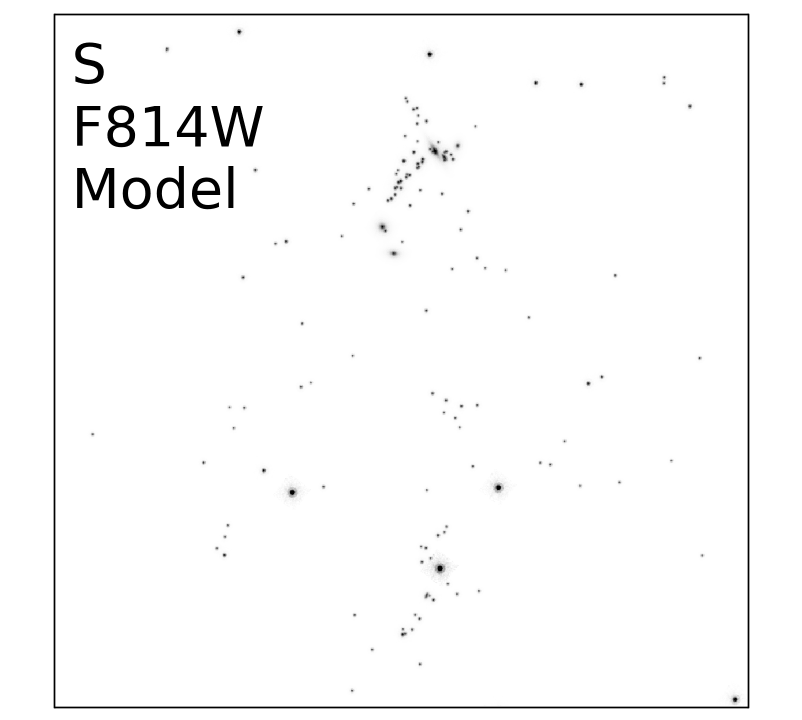}
\includegraphics[width=5.0cm]{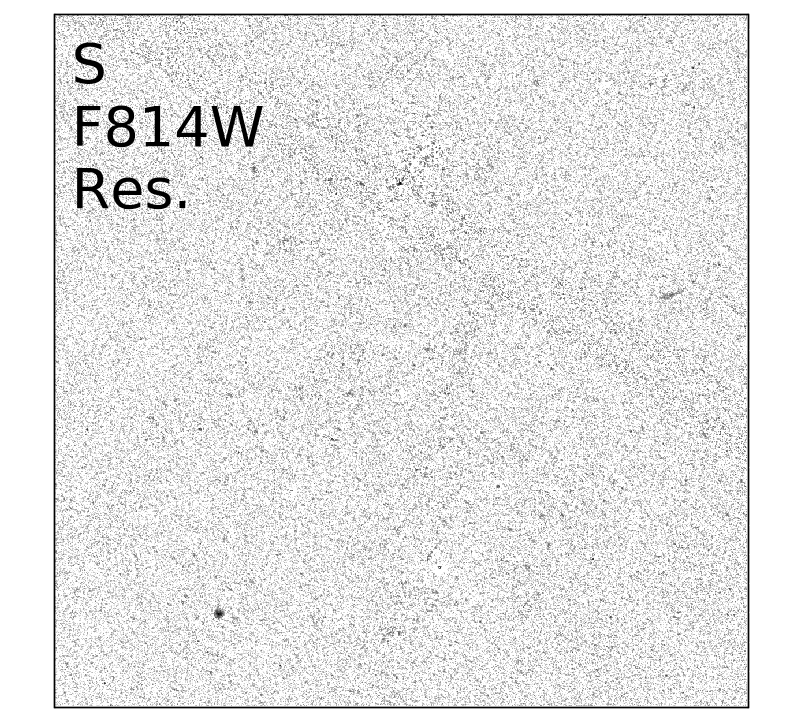}\\
\includegraphics[width=5.0cm]{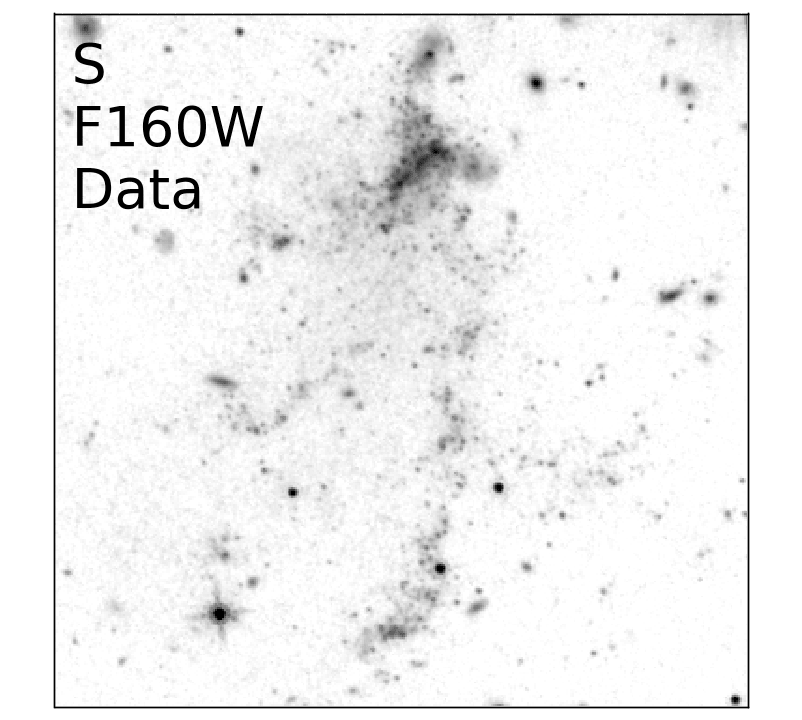}
\includegraphics[width=5.0cm]{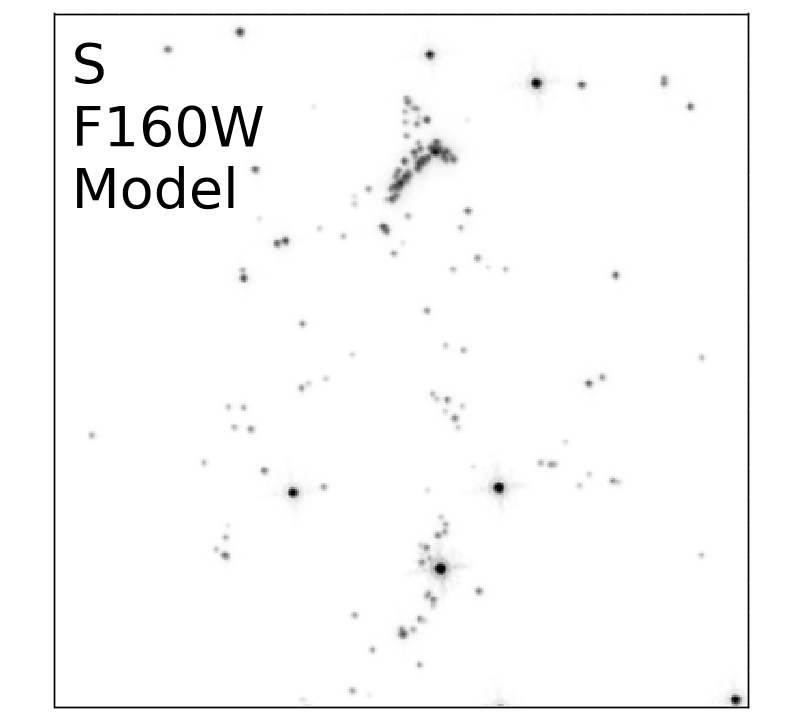}
\includegraphics[width=5.0cm]{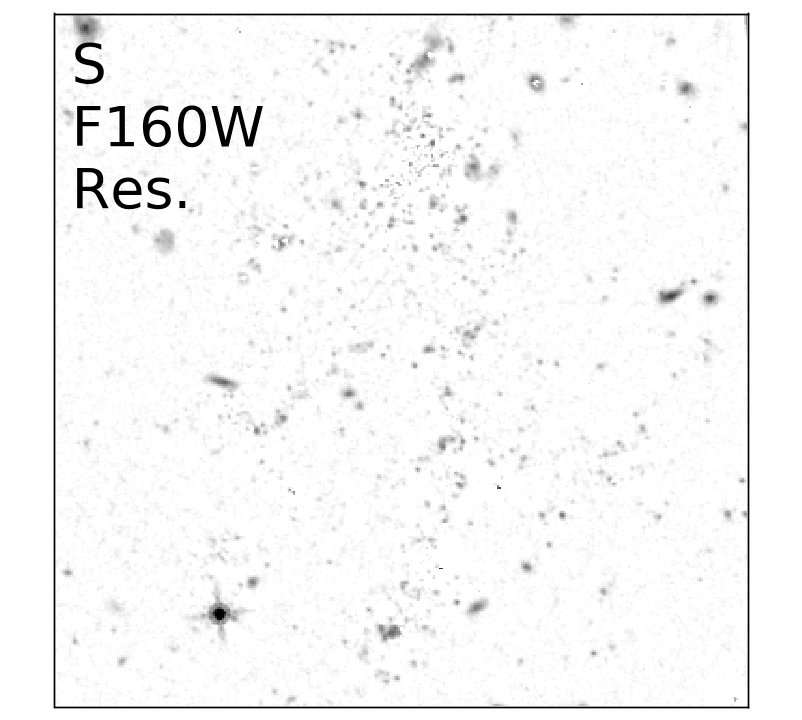}\\
\caption{Data, model and residual images for the TDG S. For each filter we show the data in the left column, the model in the middle column and the residual in the right column. From top to bottom: F336W, F475W, F606W, F814W and F160W. \label{map_S}}
\end{figure*}

\begin{figure*}
\centering
\includegraphics[width=5.0cm]{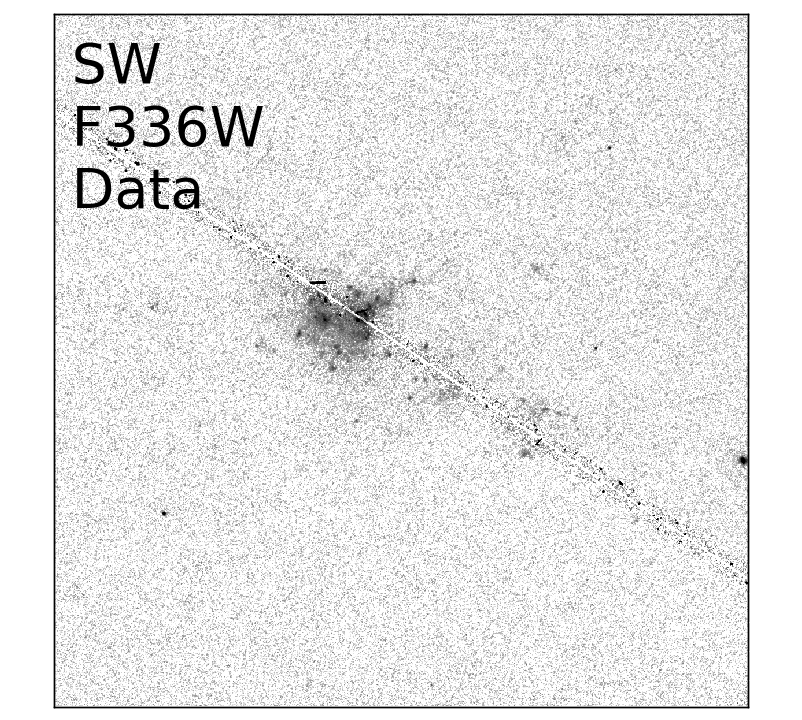}
\includegraphics[width=5.0cm]{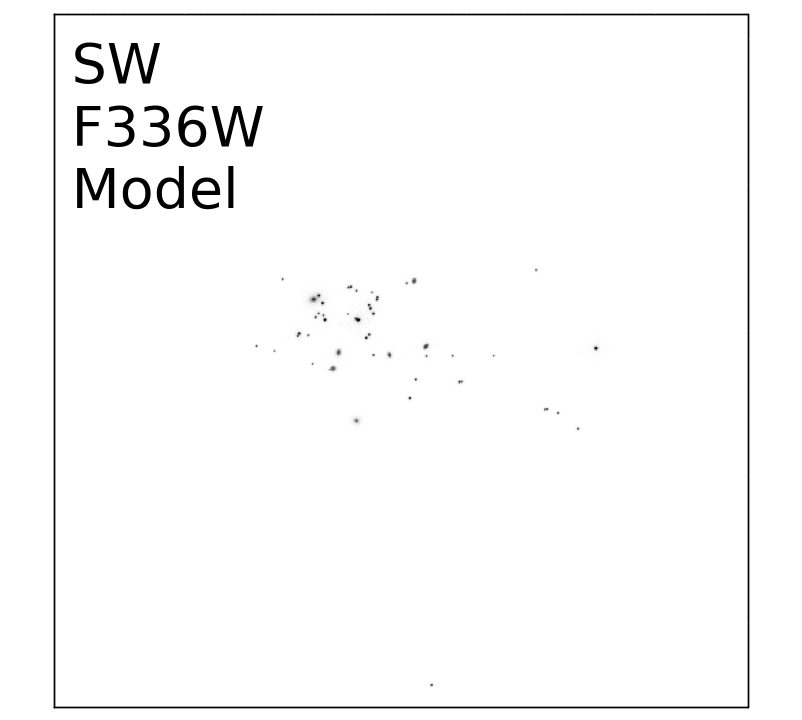}
\includegraphics[width=5.0cm]{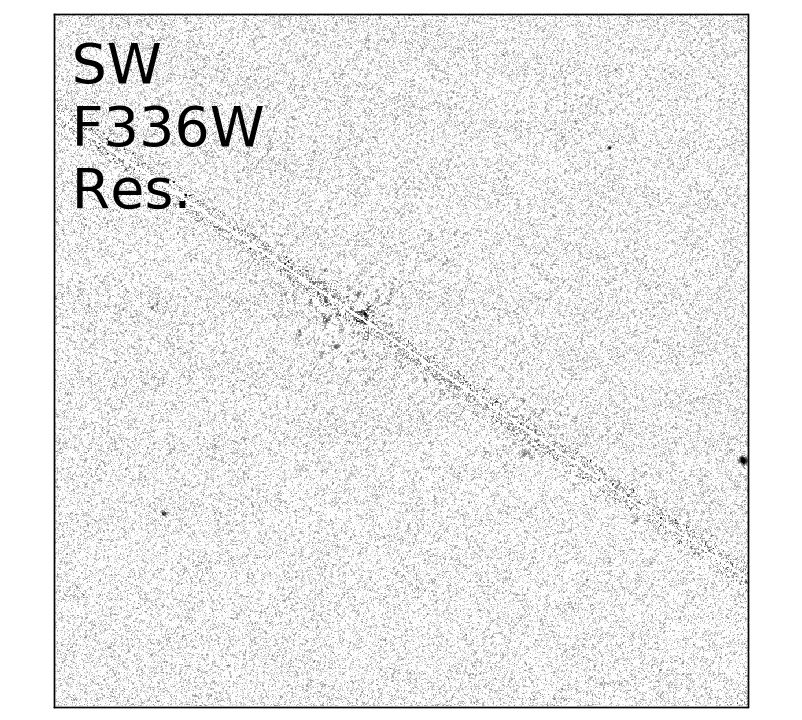}\\
\includegraphics[width=5.0cm]{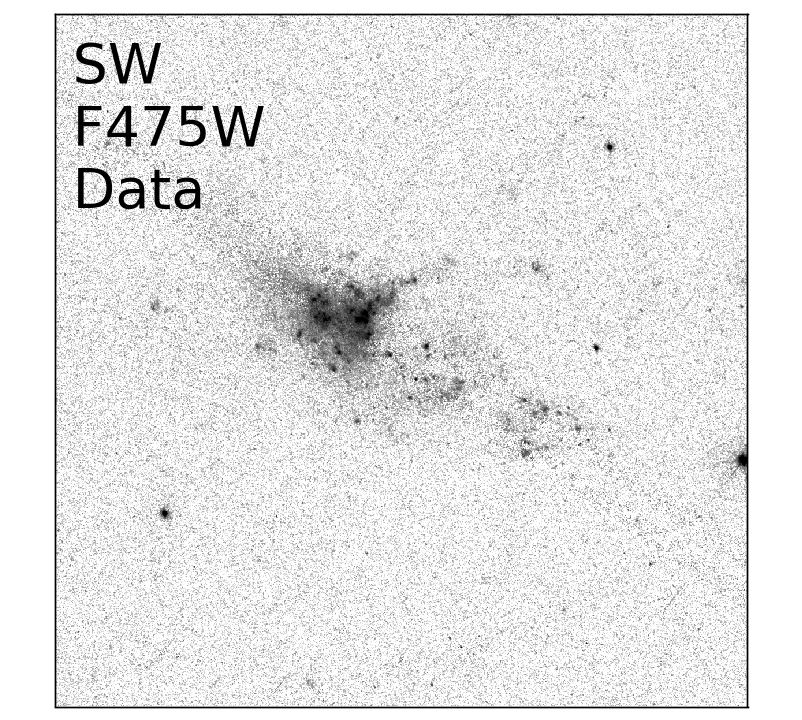}
\includegraphics[width=5.0cm]{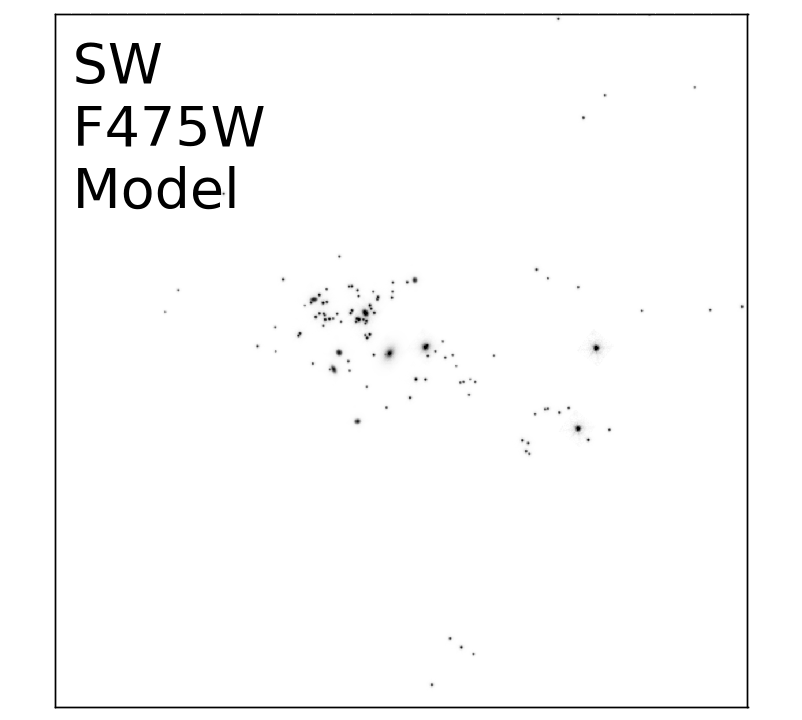}
\includegraphics[width=5.0cm]{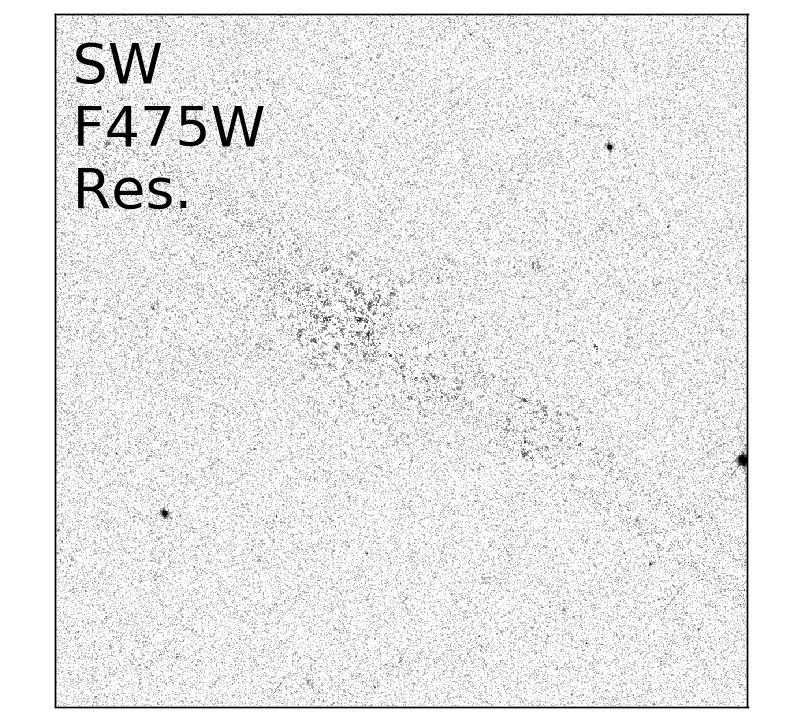}\\
\includegraphics[width=5.0cm]{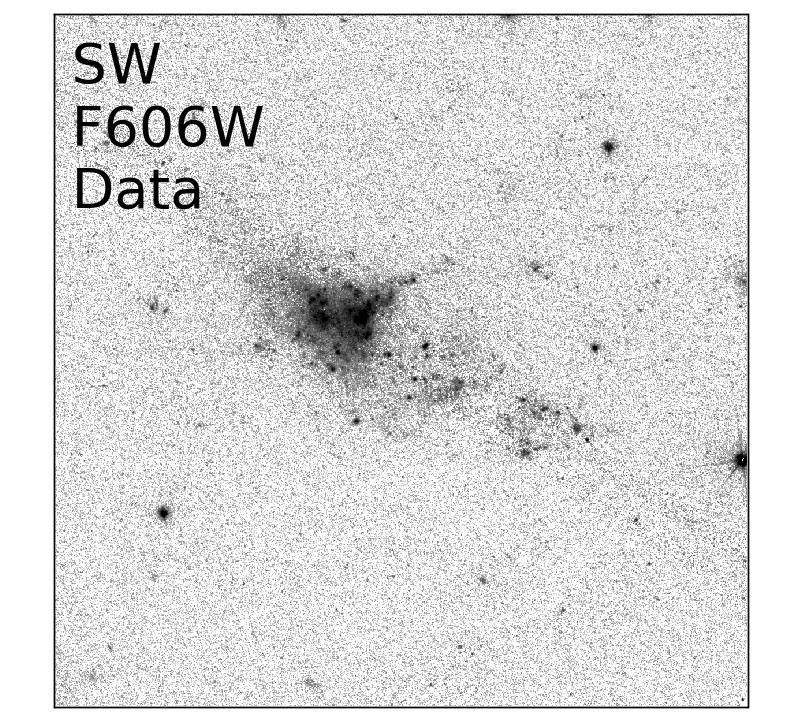}
\includegraphics[width=5.0cm]{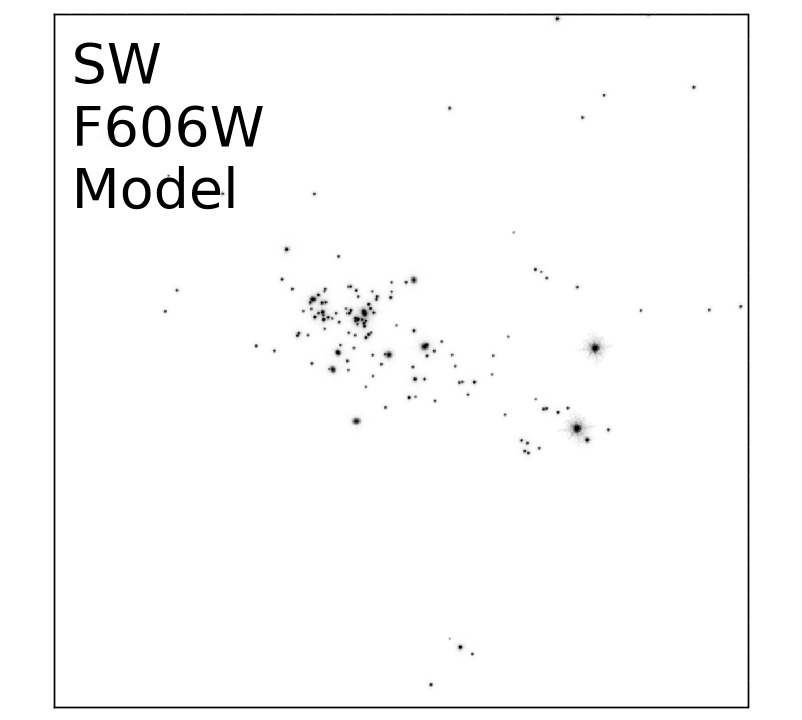}
\includegraphics[width=5.0cm]{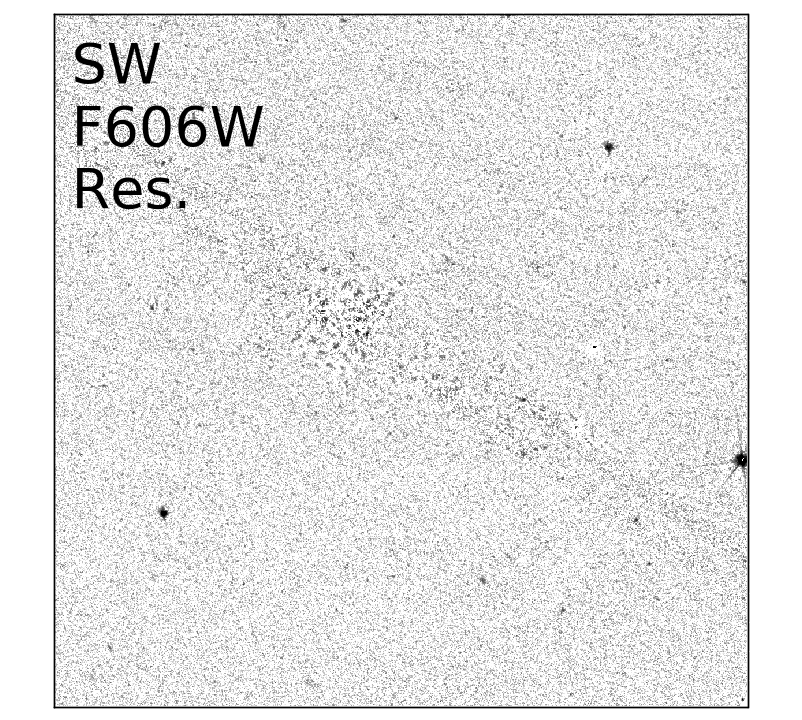}\\
\includegraphics[width=5.0cm]{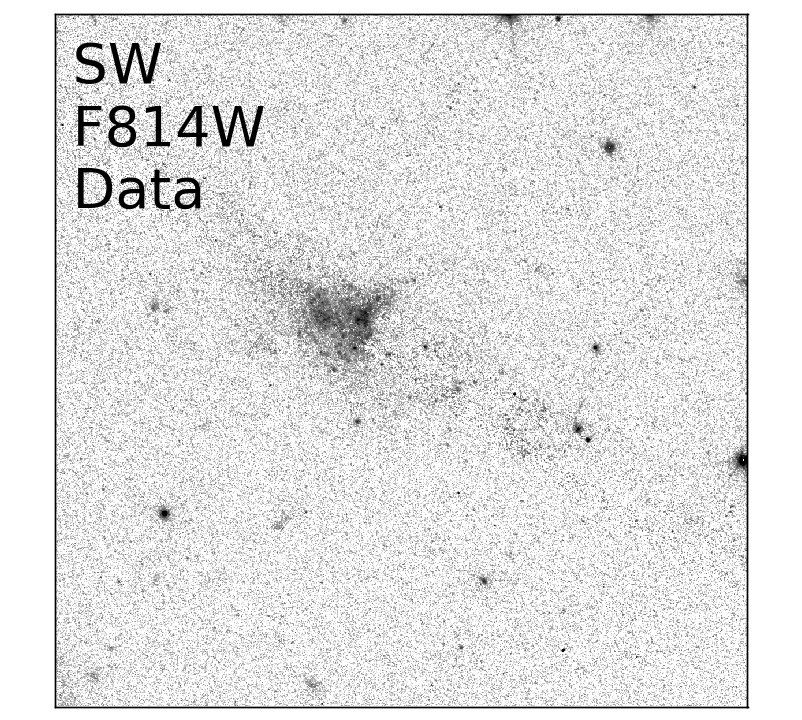}
\includegraphics[width=5.0cm]{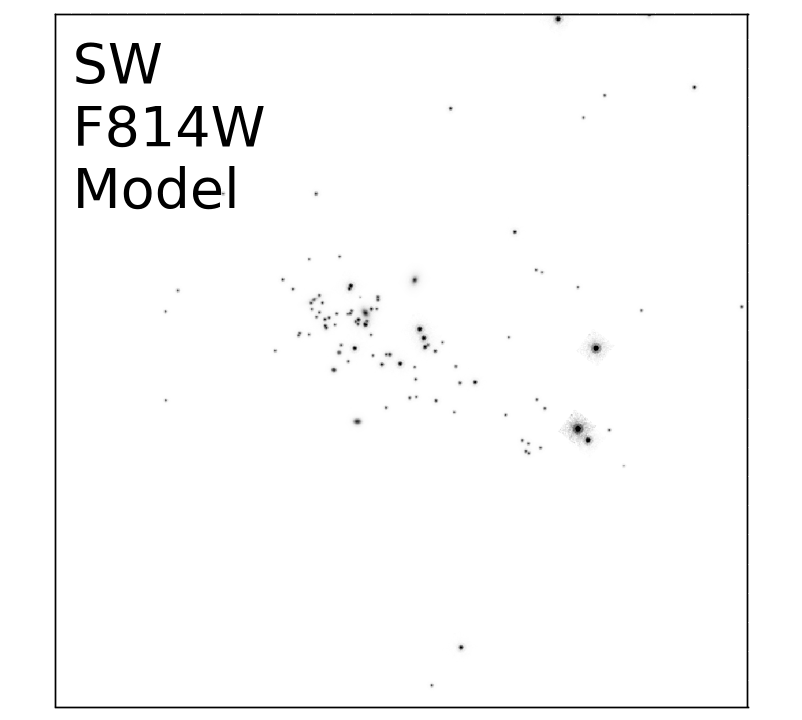}
\includegraphics[width=5.0cm]{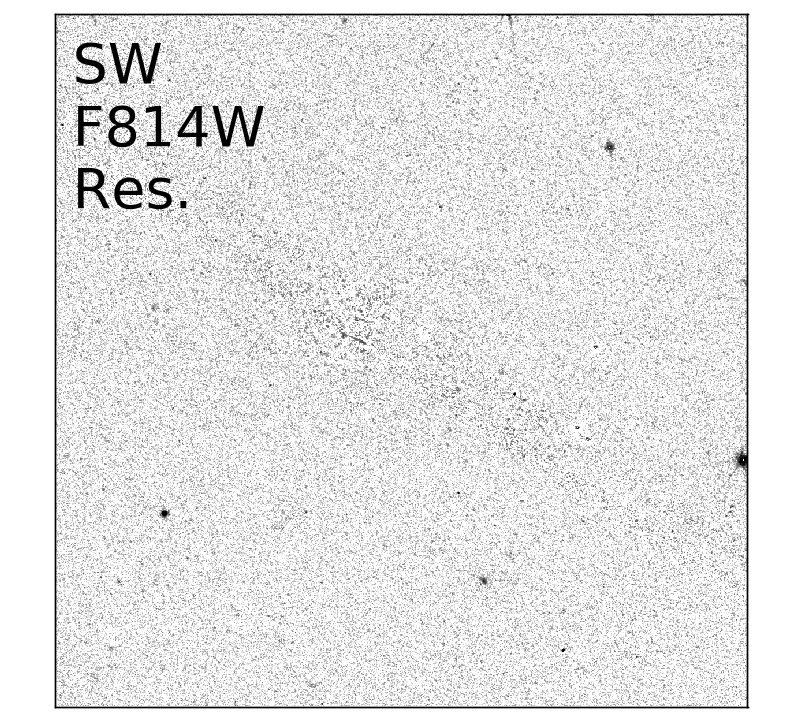}\\
\includegraphics[width=5.0cm]{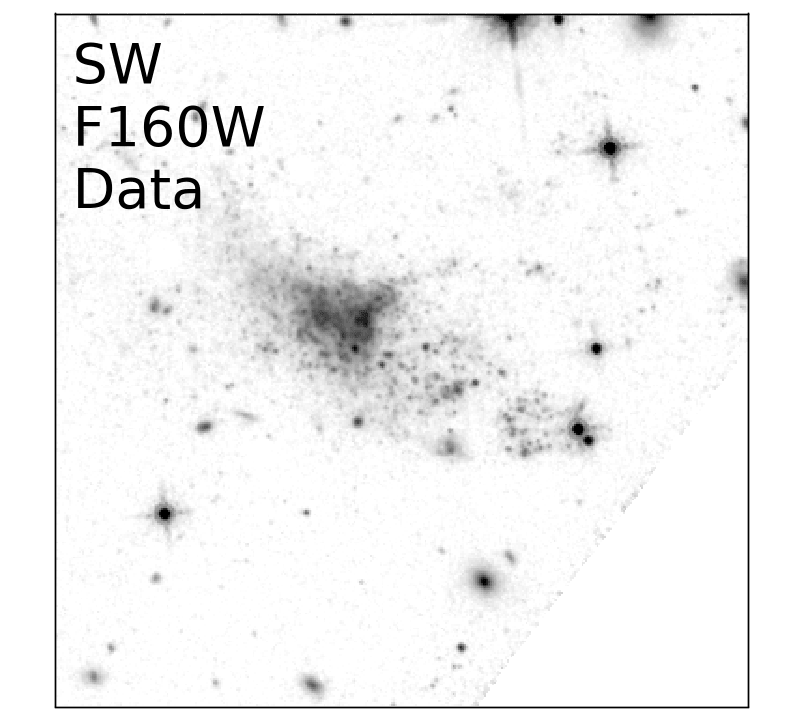}
\includegraphics[width=5.0cm]{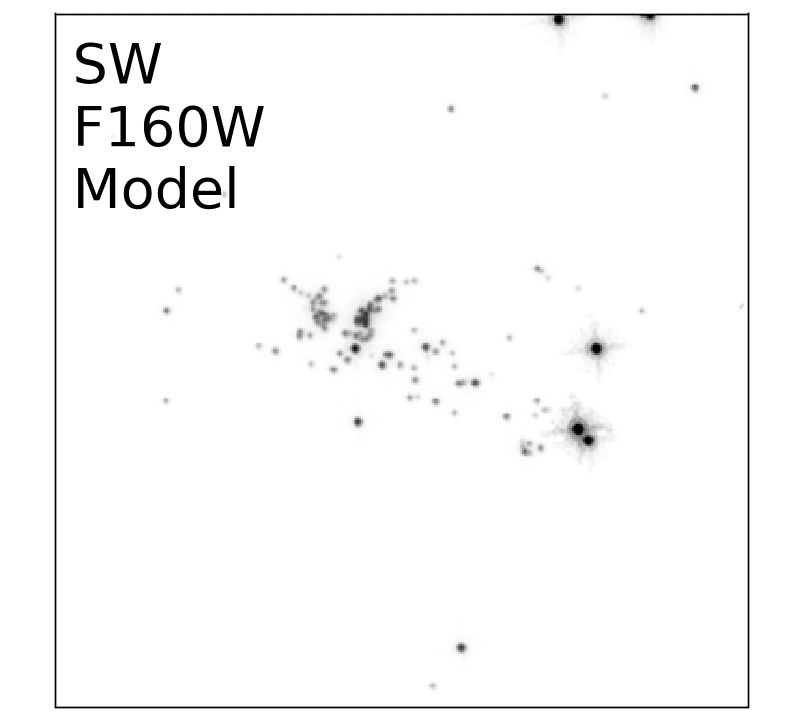}
\includegraphics[width=5.0cm]{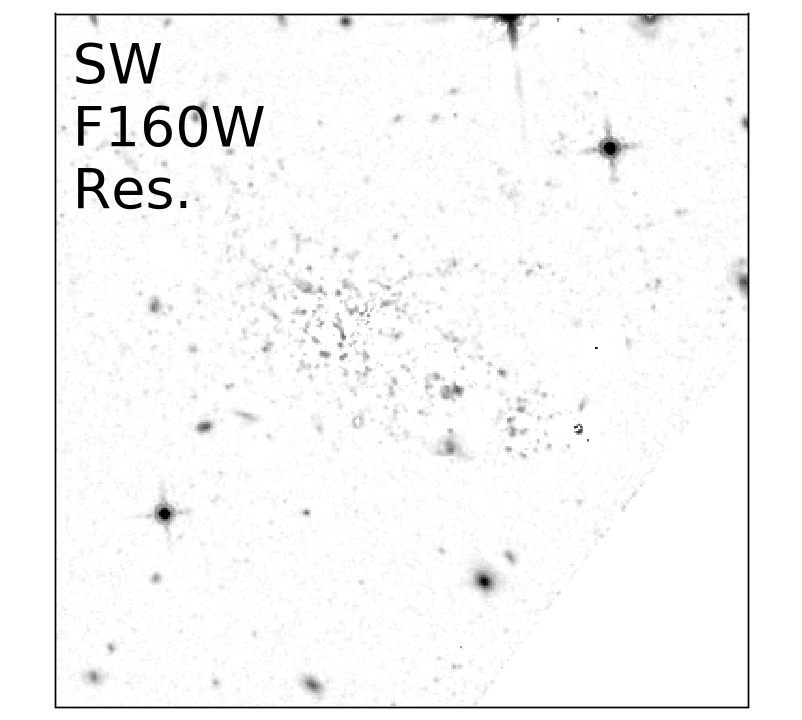}\\
\caption{Data, model and residual images for the TDG SW. For each filter we show the data in the left column, the model in the middle column and the residual in the right column. From top to bottom: F336W, F475W, F606W, F814W and F160W. \label{map_SW}}
\end{figure*}

\section{\\Sample selection}
\label{Annex_deg}

The degeneracy between age and extinction prevented us from building a complete sample. We thus used a sample which was defined using the age PDF output from CIGALE. In the following we consider two other samples.

\begin{itemize}

\item{\emph{Secure} sample: the retrieved age is younger than 30~Myr and $\mathrm{P}[  \mathrm{age} < 40] >  0.9$. }
\item{\emph{Fiducial} sample: the mode of the age PDF is below 30~Myr and $\mathrm{P}[  \mathrm{age} < 40] >  0.5$.This sample is the one used in our main study.}
\item{\emph{Inclusive} sample: $\mathrm{P}[ 0 < \mathrm{age} < 40]  >  0.1$ and we use the mass obtained by using as new age prior [1~Myr, 30~Myr]. }
\end{itemize}

On the one hand, the \emph{Secure} sample only includes clusters which are securely younger than 30~Myr, but will miss all clusters affected by degeneracies and clusters with age PDFs that are not narrow enough. On the other hand, the \emph{Inclusive} sample includes most clusters which are younger than 30~Myr, but will include a significant number of older clusters which may ressemble young clusters in our photometry. The \emph{Secure} sample is a subset of our fiducial study which is a subset of the \emph{Inclusive} sample. These two samples may thus provide us with lower or upper limits. \\

The CMFs obtained for these three samples are shown in Fig.~\ref{an_CMF}. We see that the \emph{Secure} and \emph{Inclusive} samples have a shallower and a steeper mass distribution respectively.\\

The CFEs inferred from these two samples are summarised in Table~\ref{CFE_ap} and shown in Fig.~\ref{an_CFE}. The CFE for the \emph{Secure} (resp. \emph{Inclusive}) sample can be considered as lower (resp. upper) limits to the CFE of the genuine sample of clusters younger than 30~Myr. The offset between the data and the relations from the literature is given in Table~\ref{sigma_ap}. Considering the \emph{Secure} sample, the group defined by TDG N, SW and S is respectively 2.8$\sigma$, 1.8$\sigma$ and 1.1$\sigma$ above the K12, J16 and C17 relations. Considering the \emph{Inclusive} sample, the group defined by TDG N, SW and S is respectively 4.3$\sigma$, 3.7$\sigma$ and 3.3$\sigma$ above the K12, J16 and C17 relations.

\begin{figure}
\centering
\includegraphics[width=10cm]{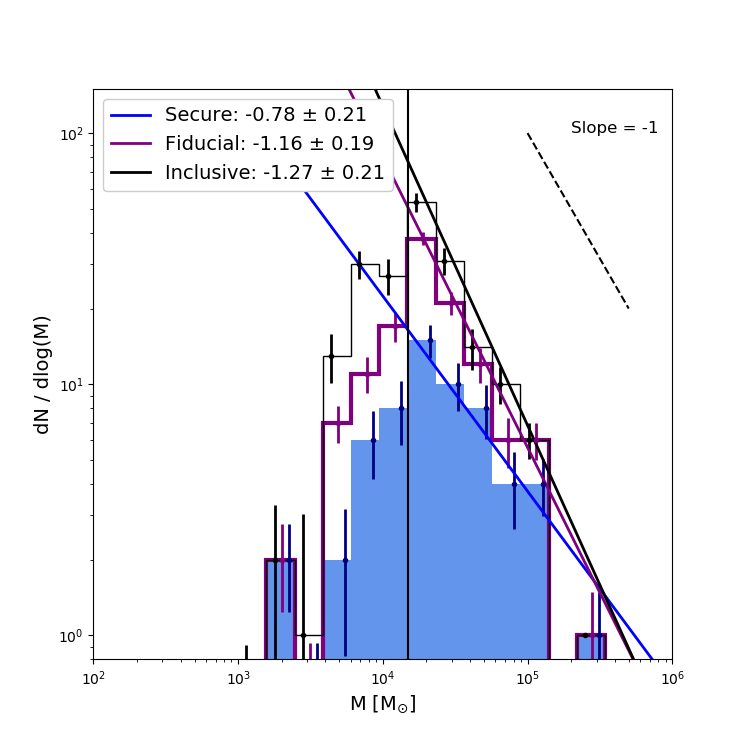}
\caption{Same legend as Fig.~\ref{CMF}. The blue (resp. black) histogram and line show the mass histogram and fit for the Secure (resp. Inclusive) samples. \label{an_CMF}}
\end{figure}

\begin{figure}
\includegraphics[width=9cm]{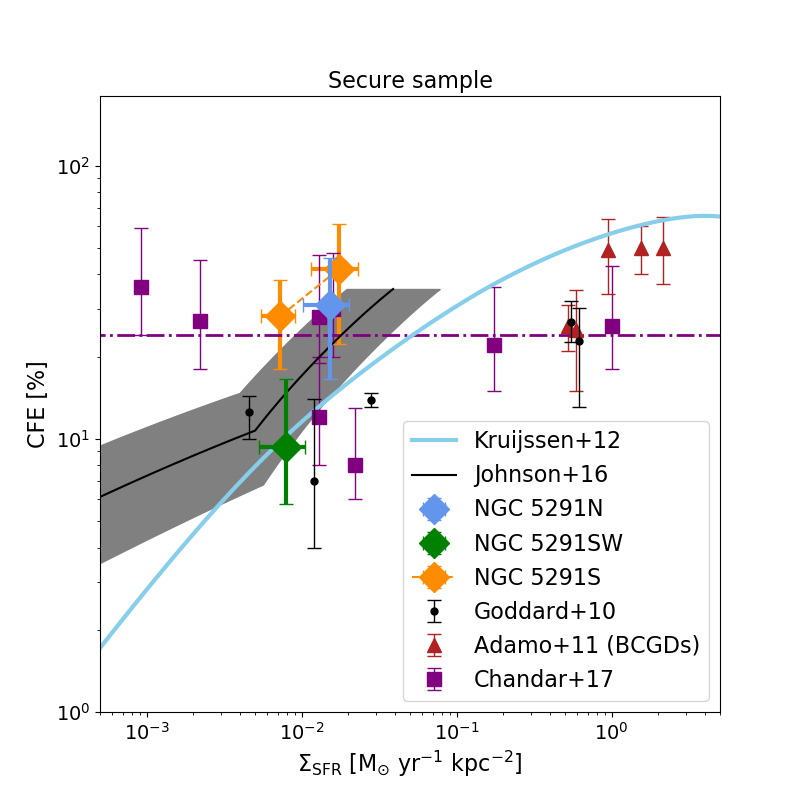}
\includegraphics[width=9cm]{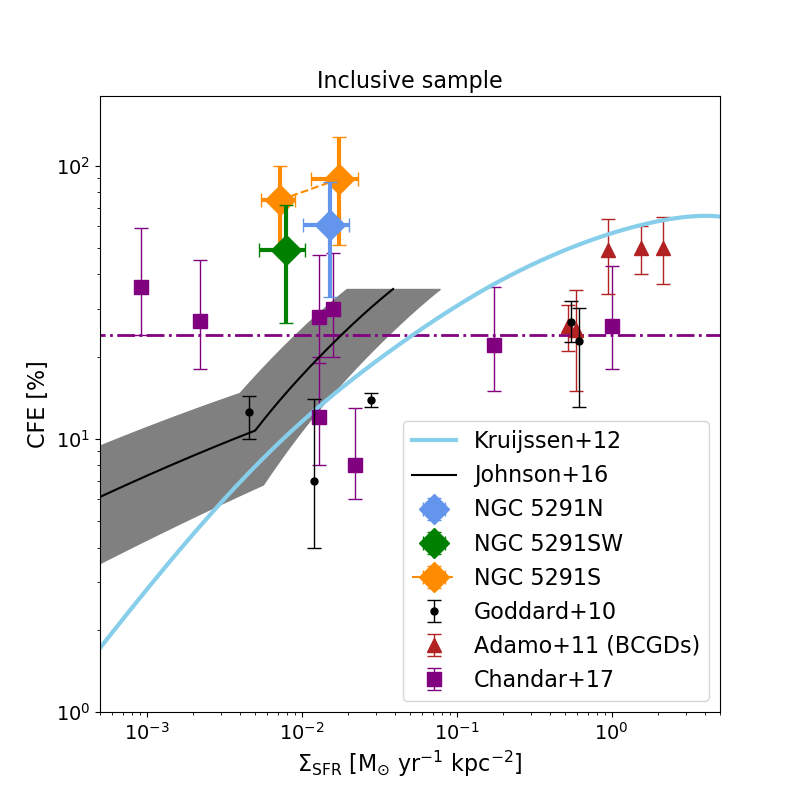}
\caption{Same legend as Fig.~\ref{gamma}. \label{an_CFE}}
\end{figure}

\begin{table}[]
\centering
\caption{Values of the CFE for the three TDGs for the three age samples considered (see text).  \label{CFE_ap}}
\label{my-label}
\begin{tabular}{l c  c} \hline\hline
      & CFE [$\%$]    & CFE [$\%$]    \\ 
Galaxy             &   \it{Secure}    &   \it{Inclusive} \\    \hline  
TDG N & 30$^{+15}_{-14}$  & 60$^{+26}_{-27}$ \\ 
TDG SW & 9$^{+7}_{-3}$  & 48$^{+23}_{-22}$ \\ 
TDG S  & 28$^{+10}_{-10}$&  74$^{+25}_{-25}$ \\
including S* & 41$^{+18}_{-19}$ &  89$^{+38}_{-38}$\\ \hline
\end{tabular}
\end{table}

\begin{table}[]
\centering
\caption{Significance, in standard deviation, of the offset between the TDGs and the relations from the literature.  \label{sigma_ap}}
\label{my-label}
\begin{tabular}{l c c c} \hline\hline
Galaxy             &   K12  &  J16  &  C17 \\    \hline  
\it{Secure} sample & & \\
TDG N &  1.1 & 0.4 & 0.3\\ 
TDG SW & 0.0 & 0.0 & 0.0 \\
TDG S  &  1.8 & 1.1 & 0.3 \\
including S* & 1.4 & 0.9 & 0.8 \\ 
TDG N+SW+S & 2.8 & 1.8 & 1.1  \\
TDG N+SW+S* & 2.5 & 1.7 & 1.5 \\ \hline 
\it{Inclusive} sample & & \\
TDG N & 1.8 & 1.4 & 1.3  \\
TDG SW & 1.6 & 1.4 & 1.0 \\
TDG S  &  2.6 & 2.3 & 1.9 \\
including S* & 2.0 & 1.7 & 1.7  \\ 
TDG N+SW+S & 4.3 & 3.7 & 3.3  \\
TDG N+SW+S* & 3.9 & 3.4 & 3.2 \\ \hline 
\end{tabular}
\end{table}

\end{document}